\documentclass[apj,twocolumn,twocolappendix,numberedappendix]{openjournal}
\usepackage{newtxtext,newtxmath,enumitem,multirow,ctable}
\usepackage[T1]{fontenc}
\usepackage[breaklinks,colorlinks,citecolor=blue,urlcolor=blue]{hyperref}

\newcommand{\Alf}{{Alfv\'en}}

\newcommand{\orcidauthor}[3]{\author{\href{http://orcid.org/#1}{#2$^{#3}$}}}

\shorttitle{A CR Origin of Extended X-ray Halos}
\shortauthors{Hopkins et al.}

\begin{document}

\title{\vspace{-0.8cm}Cosmic Rays Masquerading as Hot CGM Gas: \\
An Inverse-Compton Origin for Diffuse X-ray Emission in the Circumgalactic Medium\vspace{-1.5cm}}

\email{* phopkins@caltech.edu}
\email{** quataert@princeton.edu}

\orcidauthor{0000-0003-3729-1684}{Philip F. Hopkins}{1 *}
\orcidauthor{0000-0001-9185-5044}{Eliot Quataert}{2 **}
\orcidauthor{0000-0002-7484-2695}{Sam B. Ponnada}{1}
\orcidauthor{0000-0002-1616-5649}{Emily Silich}{1}
\affiliation{$^{1}$TAPIR, Mailcode 350-17, California Institute of Technology, Pasadena, CA 91125, USA}
\affiliation{$^{2}$Department of Astrophysical Sciences, Princeton University, Princeton, NJ 08544, USA}

\begin{abstract}
Observations from {\em ROSAT} and {\em eROSITA} have argued that Milky Way (MW), Andromeda, and lower-mass galaxies exhibit extended soft  X-ray ($1\,$keV) diffuse halos out to radii $R\gtrsim 100\,$kpc in the circumgalactic medium (CGM). If interpreted as thermal emission from hot gas, the surprisingly shallow surface brightness profiles $S_{X} \propto R^{-1}$ of this emission are difficult to explain, and contradict other observations.  
We show that such halos instead arise from inverse Compton (IC) scattering of CMB photons with GeV cosmic ray (CR) electrons. 
GeV electrons have long ($\sim$\,Gyr) lifetimes and escape the galaxy, forming a shallow extended radial profile out to $R\gtrsim 100\,$kpc, where IC off the CMB dominates their losses and should produce soft, thermal-like X-ray spectra peaked at $\sim 1\,$keV. 
The observed keV halo luminosities and brightness profiles agree well with those expected for CRs observed in the local interstellar medium (LISM) escaping the galaxy, with energetics consistent with known CRs from SNe and/or AGN, around galaxies with stellar masses $M_{\ast} \lesssim 2\times 10^{11}\,M_{\odot}$. 
At higher masses observed X-ray luminosities are larger than predicted from IC and should be dominated by hot gas. 
In the MW+M31, the same models of escaping CRs reproduce {\em Fermi} $\gamma$-ray observations if we assume an LISM-like proton-to-electron ratio and CR-pressure-dominated halo. In all other halos, the associated non-thermal radio and $\gamma$-ray brightness is far below detectable limits. 
If we have indeed detected the expected IC X-ray halos, the observations provide qualitatively new and stringent constraints on the properties of the CGM and CR physics:  the observed X-ray brightness {\em directly} traces the CR lepton energy density $e_{\rm cr,\,\ell}$ in the CGM (without any degenerate parameters).  The implied $e_{\rm cr,\,\ell}$ agrees well with LISM values at radii $R\lesssim 10\,$kpc, while following the profile predicted by simple steady state models of escaping CRs at larger radii. 
The inferred CR pressure is a major part of the total pressure budget in the CGM of Milky Way-mass galaxies, suggesting that models of thermally dominated halos at Milky Way mass may need to be revised. The measurement of X-ray surface brightness and total luminosity allows one to further determine the effective CGM diffusivity/CR streaming speed at radii $\sim 10-1000\,$kpc.  We show these also agree with LISM values at small radii but the inferred diffusivity increases significantly at larger radii, consistent with independent CGM constraints from UV absorption at $\sim 100\,$kpc.  
\end{abstract}


\maketitle

\section{Introduction}
\label{sec:intro}

Soft X-ray ($\sim 0.5-2\,$keV) observations from eROSITA (\citealt{luskova:2023.erosita.xray.cluster.surface.brightness.profiles,zhang:2024.hot.cgm.around.lstar.galaxies.xray.surface.brightness.profiles,zhang:2024.erosita.hot.cgm.around.lstar.galaxies.detected.and.scaling.relations,bahar:2024.agn.feedback.in.groups.constraints.and.tests.from.erosita}; consistent with prior but less-sensitive ROSAT observations \citep{anderson:2013.rosat.extended.cgm.xray.halos,anderson:2015.rosat.xray.halo.stacking.scaling.relations}) 
have argued that stacked X-ray images around galaxies with stellar masses $\sim 10^{10}-10^{11.5}\,M_{\odot}$, including Milky Way (MW) and Andromeda (M31)-mass and lower-mass galaxies, exhibit an excess of diffuse emission that appears to be CGM emission from $\gtrsim 10-100\,$kpc around the central galaxies. 
Surprisingly, the X-ray surface brightness or flux  declines very slowly, $\propto 1/R$ with galacto-centric distance even approaching $\gtrsim R_{500}$ -- much shallower than the observed X-ray halos around more massive, individually-detected groups and clusters \citep{luskova:2023.erosita.xray.cluster.surface.brightness.profiles}. The emission is also much more luminous, and declines much more slowly with radius, than predicted in high-resolution numerical simulations of MW-M31-mass systems, if it is assumed to arise from thermal (free-free+metal-line) hot gas emission (\citealt{truong:2023.cosmo.sims.sb.predictions.verylow.larger.vs.erosita.data,sultan:2024.cooling.flows.model.fire.hot.cgm.lowermass.galaxies,popesso:2024.erosita.stacking.lx.mhalo.by.halo.mass}; Silich et al., in prep.). And regardless of models, reproducing the observed emission via hot gas may require more baryons (uniformly at {\em super-virial} temperatures) than the universal baryon fraction, or more metals than produced by the sum of all SNe in the history of the galaxy, or even more radical cosmological revisions, for MW-mass systems (see \S \ref{sec:thermal} below and \citealt{zhang:2024.hot.cgm.around.lstar.galaxies.xray.surface.brightness.profiles,lau:2024.erosita.profiles.require.cosmological.constraints.violations}), as well as directly contradicting other observational constraints from both UV \citep{werk:2014.cos.halos.cgm,faerman:2022.cgm.props.needed.obs.vs.sams,wijers:2024.neviii.failure.of.simulations} and X-ray {\em absorption}-line studies from both eROSITA \citep{ponti:2023.erosita.supervirial.gas.close.to.galaxy.cgm.low.metal.and.low.density} and Chandra \citep{yao:2010.chandra.upper.limits.warm.hot.cgm.gas.in.mw.mass.halos}. 
Both eROSITA and ROSAT also find a sharp change in behavior in the total halo luminosity $L_{X}$ versus stellar mass $M_{\ast}$, where below $M_{\ast} \lesssim 2\times 10^{11}\,M_{\odot}$ the correlation is shallow (linear or sub-linear), while it steepens dramatically at higher $M_{\ast}$. 

These observations suggest that there is either a dramatic change in halo gas properties in lower mass systems or a change in the nature of the dominant soft X-ray emission process in lower-mass halos. In this paper, we show that inverse Compton (IC) scattering of CMB photons with $\sim $\,GeV CR electrons (which are observed to escape the ISM), naturally predicts X-ray halos very similar to those observed in lower-mass systems.

It has long been known that CRs can up-scatter CMB photons into X-ray bands \citep{blumenthal:1970.cr.loss.processes.leptons.dilute.gases,raphaeli:1979.cluster.xray.emission.cr.ic.vs.thermal,sarazin:1988.book.cluster.xray.emission,sarazin:1999.cr.electron.emission.cluster.centers.xrays}, and many previous studies have argued this could be important for observed X-ray emission in the centers of some galaxy clusters \citep[e.g.][]{hwang:1973.cluster.center.emission.in.euv.from.cr.ic,sarazin:1999.cr.electron.emission.cluster.centers.xrays,gitti:2002.perseus.minihalo.xray.inverse.compton,gitti:2004.minihalo.abell.2626.reaccel,bonamente:2007.abell.3112.clear.xray.inverse.compton.required.luminosity.fits.models.gamma.rays.too,murgia:2010.ophiuchus.cluster.minihalo.xray.inverse.compton,bartels:2015.radio.inverse.compton.cluster.minihalo.prospects,gitti:2016.radio.minihalos.coolcore.clusters.candidates.review}, as well as specific regions in the MW (e.g.\ \citealt{porter:2006.ic.galactic.snrs.isrf.effects,porter:2008.galactic.ridge.cr.ic.xray.gamma.ray.emission.multiprocess,strong:2010.galprop.multiwavelength.cr.emission.across.wide.range}, and for higher-energy emission, \citealt{ackermann:2012.fermi.gamma.rays.ism.cr.emission.modeling,orlando:2015.multi.wavelength.CR.lepton.constraints.gamma.rays.mostly,wehahn:2021.gamma.rays,werhahn:2021.cr.calorimetry.simulated.galaxies}). 
Meanwhile, a wealth of direct CR observations from experiments like Voyager, AMS-02, and others, show that near the solar circle, $\sim$\,GeV CR electrons dominate the CR lepton energy budget, with an energy density $\sim 0.02\,{\rm eV\,cm^{-3}}$, and that they have long lifetimes ($\sim 10^{9}$\,yr) compared to their residence/escape timescale from the ISM ($\lesssim 10^{7}\,$yr). This implies that GeV leptons must be escaping from the galaxy into the CGM with an implied galaxy-integrated leptonic rate $\sim 10^{40}\,{\rm erg\,s^{-1}}$ \citep[e.g.][and references therein]{zweibel:cr.feedback.review,2018AdSpR..62.2731A,evoli:2019.cr.fitting.galprop.update.ams02,maurin:2018.cr.sam.favored.parameters.close.to.fire,butsky:2022.cr.kappa.lower.limits.cgm,dimauro:2023.cr.diff.constraints.updated.galprop.very.similar.our.models.but.lots.of.interp.re.selfconfinement.that.doesnt.mathematically.work}. In this paper, we show that these escaping CR leptons produce X-ray halos\footnote{Throughout, we use the term ``halo'' in the same sense as the eROSITA observations compared and CGM community, referring to gas out to the virial radius $\sim 300\,$kpc from the galaxy center. But note in the CR community, ``halo'' typically refers to the ISM thick disk or disk-coronal region extending just a few kpc above the star-forming thin disk (much smaller than our focus).} with a spectrum peaked in the observed soft X-ray bands, and with a normalization, luminosity, and surface brightness profile very similar to those observed around galaxies with stellar masses $M_{\ast} \lesssim 2\times 10^{11}\,M_{\odot}$. 

If these observed X-ray halos indeed arise from IC, then they do not trace hot gas, but instead trace the properties of CRs. This enables, for the first time, direct measurement of the CR (leptonic) energy density in the CGM, as well as the CR transport parameters in the CGM -- profoundly powerful and unique constraints for CR ``feedback'' and scattering physics models.

We stress that what is new here is not the suggestion that CR-IC can produce X-ray emission (this is well-known). Rather, what has not been appreciated in the CGM community studying $\sim L^{\ast}$ galaxies (as evidenced in the literature cited above and in \S~\ref{sec:thermal}), which we argue here, is that (1) CR-IC could be the dominant source of soft X-rays at {\em large} radii (of order the virial radius) in halos; (2) this is especially relevant in lower halos masses (specifically MW-M31 mass and below), given the observed scaling of star formation rates (SFRs) and stellar masses $M_{\ast}$ on the SF ``main sequence,''  $M_{\ast}-M_{\rm halo}$, and $L_{X}-M_{\ast}$ relations; and (3) this CR-IC emission may well have been observed already, but has been widely interpreted as thermal emission.

\section{Inverse Compton from Extended Cosmic Ray Halos}
\label{sec:IC}

\subsection{Basic Scalings of Emission and Surface Brightness Profiles}
\label{sec:basic}

Consider the inverse Compton (IC) radiation from a population of CR electrons. A typical CR electron with energy $E_{\rm cr} \sim E_{\rm cr, GeV}\,{\rm GeV}$ (around the peak of the CR spectrum at $E_{\rm cr, GeV} \sim 1$) will IC scatter a photon to a characteristic energy $h \nu_{\rm IC} \rightarrow h \nu_{\rm initial} \,\gamma_{\rm cr}^{2}$ \citep{blumenthal:1970.cr.loss.processes.leptons.dilute.gases,rybicki.lightman:1979.book}, where for CMB photons $\langle h \nu_{\rm initial} \rangle \sim 3\,k_{B} T_{\rm cmb} \sim 7\times 10^{-4} (1+z)\,{\rm eV}$, so at emission (rest-frame) 
\begin{align}
\label{eqn:peak} h \nu_{\rm IC} \sim1-3\, {\rm keV}\,E_{\rm cr, GeV}^{2}\,(1+z)\ .
\end{align}
The X-ray emissivity will be given by \citep{rybicki.lightman:1979.book}
\begin{align}
\label{eqn:emissivity.oom} \epsilon_{X} &\sim \frac{4}{3}\,h \nu_{\rm IC}\,n_{{\rm cr},\,e^{\pm}}\,n_{\rm photons}\,\sigma_{\rm T}\,c \\ 
\nonumber &\sim 10^{34.5}\,\frac{\rm erg}{\rm s\,kpc^{3}}\left( \frac{e_{\rm cr,\,\ell} E_{\rm cr, GeV} }{\rm 0.02\,eV\,cm^{-3}}\right) \,(1+z)^{4}
\end{align}
where $e_{\rm cr,\,\ell}$ is the total {\em leptonic} CR energy density and we have scaled to the Solar-neighborhood value at $\sim$\,GeV.\footnote{We can also define the ratio of leptonic-to-total CR energy, $f_{\rm cr,\,\ell} \equiv e_{\rm cr,\,\ell}/e_{\rm cr,\,{\rm tot}}$, with $f_{\rm cr,\,\ell}^{\odot} \sim 0.02$ in the LISM.}
A proper calculation of the X-ray spectrum and emissivity (akin to those in \citealt{sarazin:1999.cr.electron.emission.cluster.centers.xrays,porter:2006.ic.galactic.snrs.isrf.effects}, including higher-order terms like Klein-Nishina corrections, though these are small) are shown in Fig.~\ref{fig:spectrum} (and \S~\ref{sec:detailed} below). 
Integrating through a halo\footnote{For power-law profiles $e_{\rm cr} \propto r^{-\alpha}$, at impact parameter $R$, $S_{X} \approx A_{0}\, R\,\epsilon_{X}(r=R)$ with $A_{0}\equiv \pi^{1/2} \Gamma[(\alpha-1)/2]/\Gamma[\alpha/2] \sim 3-5$ for the range of slopes of interest here.} at some impact parameter $R \equiv R_{100}\,100\,{\rm kpc}$, this gives a surface brightness\footnote{We follow \citealt{zhang:2024.erosita.hot.cgm.around.lstar.galaxies.detected.and.scaling.relations}'s convention (since we compare to their observations) for the definition of surface brightness as $dL_{\rm iso}/dA$ in terms of the differential isotropic luminosity and area.}
\begin{align}
\label{eqn:sb.oom} \frac{S_{X,\,{\rm keV}}}{10^{37} \rm erg\,s^{-1}\,kpc^{-2}} &\sim 
\left( \frac{e_{\rm cr,\,\ell} E_{\rm cr, GeV} }{\rm 0.02\,eV\,cm^{-3}}\right)  \left( \frac{R}{100\,{\rm kpc}} \right)\, (1+z)^{4}\ .
\end{align}

\begin{figure}
	\centering\includegraphics[width=0.99\columnwidth]{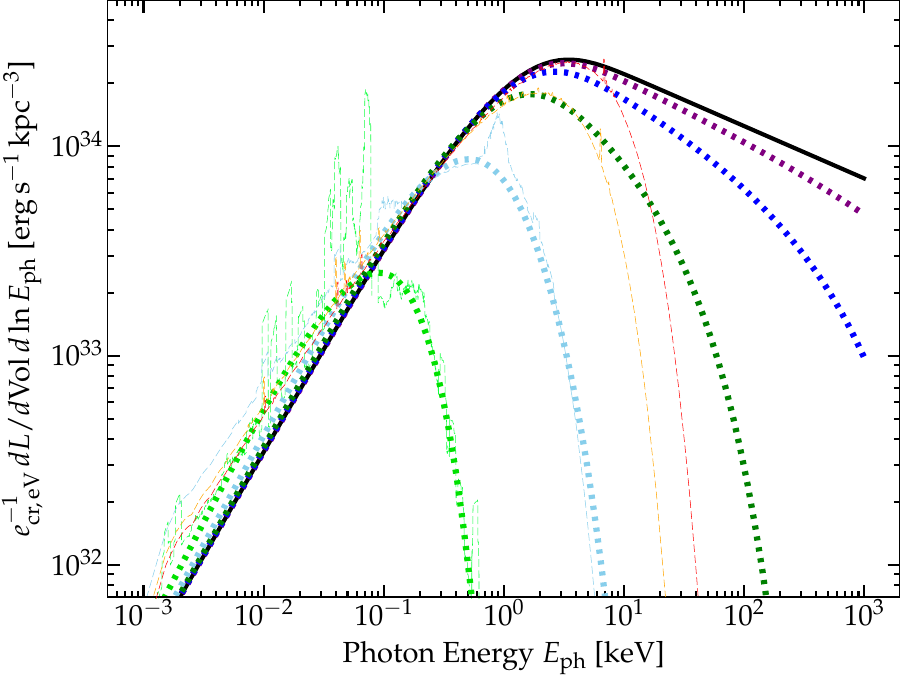} 
	\centering\includegraphics[width=0.99\columnwidth]{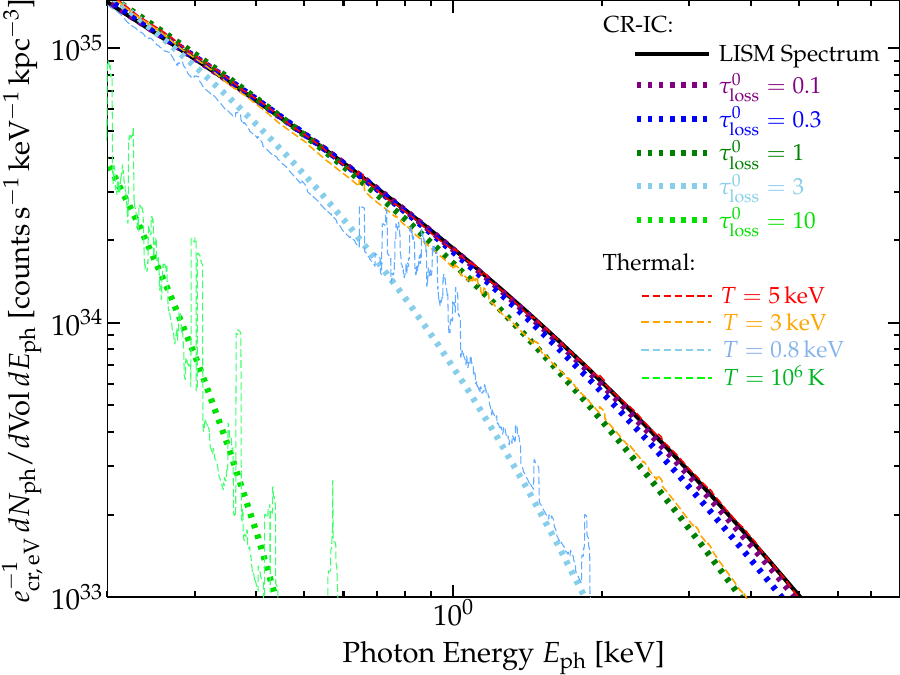} 
	\caption{Observed-frame X-ray spectra produced by IC from CRs off the CMB (\S~\ref{sec:basic} \&\ \ref{sec:detailed}).
	{\em Top:} $E^{2} dN/dE$ over a very broad energy range.
	{\em Bottom:} $dN/dE$ focused on standard X-ray energies $0.2-8$\,keV.
	We show the spectrum from IC only, from the CMB (other cosmic backgrounds are totally negligible at these energies), at $z=0$, from standard expressions.
	For the CR lepton spectrum, we compare (1) the observed LISM spectrum, and (2) that spectrum corrected for IC losses with finite travel time (i.e.\ finite distance from the ISM) parameterized by $\tau_{\rm loss}^{0} \equiv t_{\rm travel}/t_{\rm loss} \sim (R/100\,{\rm kpc})\,(100\,{\rm km\,s^{-1}}/v_{\rm st,\,eff})$ at $1\,$GeV. 
	For reference, we compare APEC thermal emission spectra (free-free+line; metallicities $\sim 0.05\,Z_{\odot}$ matching the X-ray absorption upper limits at MW virial radii; \S~\ref{sec:thermal}). 
	CRs leaking from the ISM produce a thermal-like spectrum with most of the luminosity at $\sim$\,keV (Eq.~\ref{eqn:peak}; from $\sim$\,GeV leptons, near the LISM spectral peak, with $\sim$\,Gyr lifetimes), until most of the CR energy is lost ($\tau_{\rm loss}^{0} \gg 1$, expected at $R \gtrsim 0.1-1$\,Mpc distances from the galaxy; \S~\ref{sec:losses}).
	\label{fig:spectrum}}
\end{figure}

Now consider the expectation for a CR halo around a galaxy. 
In detail, modern CR-MHD simulations predict this by explicitly evolving the anisotropic multi-moment CR transport equations, allowing for non-equilibrium behavior, dynamically evolved magnetic fields, losses which depend on the local evolved fields (like gas or radiation energy density or field strength), for arbitrary streaming rates, in a fully-cosmological context \citep[for recent reviews, see][]{hanasz:2021.cr.propagation.sims.review,owen:2023.cr.review.galaxies.feedback,ruszkowski.pfrommer:cr.review.broad.cr.physics}. 
Many of these studies \citep[e.g.][]{su:turb.crs.quench,hopkins:cr.mhd.fire2,hopkins:2020.cr.outflows.to.mpc.scales,ji:fire.cr.cgm,ji:20.virial.shocks.suppressed.cr.dominated.halos,butsky:2022.cr.kappa.lower.limits.cgm,ponnada:2023.fire.synchrotron.profiles,ponnada:2023.synch.signatures.of.cr.transport.models.fire}, as well as analytic models \citep{quataert:2021.cr.outflows.diffusion.staircase,quataert:2022.isothermal.streaming.wind.analytic.cr.wind.models} have shown that the resulting CR energy density $e_{\rm cr}$ can be approximated reasonably well by simply assuming a spherically-symmetric, steady-state-flux solution with weak losses or re-acceleration in the (low-density) CGM (since the $\sim$\,GeV CRs which dominate $e_{\rm cr}$ have small losses except in dense ISM gas and after very long timescales from inverse Compton, which we account for below, with a factor $f_{\rm loss}$), given some (time-averaged) leptonic energy injection rate $\dot{E}_{\rm cr,\,\ell}$ from the galaxy (e.g.\ from SNe shocks, AGN, fast stellar winds, pulsar wind nebulae, etc.). This gives:
\begin{align}
\label{eqn:simple.prop} e_{\rm cr,\,\ell} \sim  \frac{f_{\rm loss}\,\dot{E}_{\rm cr,\,\ell}}{4\pi\,v_{\rm st,\,eff}\,r^{2}} \approx \frac{f_{\rm loss}\,\dot{E}_{\rm cr,\,\ell}}{4\pi\,\kappa_{\rm eff}[r]\,r} 
\end{align}
where $\kappa_{\rm eff}$ and $v_{\rm st,\,eff}$ are some effective, spectrum and isotropically and position-averaged CR diffusion coefficient and/or streaming (+advection) speed in the CGM.\footnote{Note that 
\citet{hopkins:cr.transport.constraints.from.galaxies,hopkins:2020.cr.transport.model.fx.galform,hopkins:2021.sc.et.models.incompatible.obs,hopkins:2022.cr.subgrid.model,butsky:2022.cr.kappa.lower.limits.cgm} show this is remains a reasonable approximation even when the CR scattering rate (hence $\kappa_{\rm eff}$ and $v_{\rm st,\,eff}$) is a complicated and highly-variable function of the local plasma properties, provided an appropriate effective average is used. And this or very similar approximations have been widely applied in modeling CRs in the CGM \citep[see e.g.][]{wiener:cr.supersonic.streaming.deriv,tsung:2021.cr.outflows.staircase,quataert:2021.cr.outflows.diffusion.staircase,quataert:2022.isothermal.streaming.wind.analytic.cr.wind.models,owen:2023.cr.review.galaxies.feedback,ramesh:2024.tng.plus.our.subgrid.crs.very.strong.fb.fx,romano:2025.cr.halo.modeling.with.simple.subgrid.model,liang:2025.leaky.boxes.levy.flights.modeling.crs.transport}. Implicitly Eq.~\ref{eqn:simple.prop} does assume the gyro-radii ($\sim$\,au) and pitch angle scattering mean-free paths ($\sim$\,pc-to-kpc) of the ($\sim$\,GeV) CRs of interest are small compared to the global length scales of the problem ($\sim 100\,$kpc), but this is easily satisfied in the CGM.}

Theoretically, the predicted values of $\kappa_{\rm eff}$ (or $v_{\rm st,\,eff}$) in the CGM remain highly uncertain, since the physical conditions are quite distinct from the ISM and Galaxy where we have more direct CR constraints -- this is indeed why detection of CR-IC would be exciting. But to give some order-of-magnitude reference point, in the CGM at the halo masses of interest, the models above and \citet{wiener:cr.supersonic.streaming.deriv,wiener:2017.cr.streaming.winds,Rusz17,thomas:2022.self-confinement.non.eqm.dynamics} typically predict that CR transport at $\sim\,$GeV is streaming-dominated with $v_{\rm st,\,eff} \sim v_{A} \sim 30-100\,{\rm km\,s^{-1}}$ in a MW-mass halo (one to a few times the \Alf\ speed), increasing to a few $100\,{\rm km\,s^{-1}}$ in small-group mass halos ($\sim 10^{13}\,M_{\odot}$). 
For the typical SNe rate in Milky Way to Andromeda-type galaxies (assuming the standard $\sim 10\%$ of the energy per SNe goes into CRs and $\sim 2\%$ of this into leptons; \citealt{higdon:crs.accel.in.superbubbles.in.sne.ejecta.mass.dominated.regions,blasi:cr.propagation.constraints,2012JCAP...07..038C}) plus mean jet power (averaged over the whole population, since the travel time is $\sim$\,Gyr so we should average over any AGN outbursts in the last Gyr) for the population (scaled to either radio or X-ray power; see \citealt{allen:jet.bondi.power}), one expects a population-averaged $\dot{E}_{\rm cr,\,\ell} \sim 10^{39}-10^{41}\,{\rm erg\,s^{-1}}$ increasing from Milky Way-to-Andromeda mass galaxies \citep[see references in][and discussion of energetics in \S~\ref{sec:ltot} below]{su:2023.jet.quenching.criteria.vs.halo.mass.fire}. 
Inserting these into Eq.~\ref{eqn:simple.prop}, we obtain: 
\begin{align}
\label{eqn:sb.oom.model} \frac{S_{X,\,{\rm keV}}}{\rm erg\,s^{-1}\,kpc^{-2}} \sim & 10^{35.3}\,
 \left(\frac{\dot{E}_{\rm cr,\,\ell}}{10^{40}\,{\rm erg\,s^{-1}}}\right) 
\left(\frac{100\,{\rm km\,s^{-1}}}{v_{\rm st,\,eff}(R)}\right) \\
\nonumber &\times 
\left(\frac{100\,{\rm kpc}}{R}\right)\,f_{\rm loss}\,
E_{\rm cr, GeV} (1+z)^{4} \\ 
\sim & 10^{35.3} (1+z)^{4} R_{100}^{-1}\,f_{\rm loss}\,\frac{\dot{E}_{40}}{v_{100}}\ ,
\end{align}
where $\dot{E}_{40} \equiv \dot{E}_{\rm cr,\,\ell}/10^{40}\,{\rm erg\,s^{-1}}$, $v_{100} \equiv v_{\rm st,\,eff}(R) / 100\,{\rm km\,s^{-1}}$, $R_{100}\equiv R/100\,{\rm kpc}$. 
This is remarkably similar to the observed scalings of $S_{X}$ in \citet{zhang:2024.hot.cgm.around.lstar.galaxies.xray.surface.brightness.profiles}, both in normalization (for the expected parameters) and (shallow) slope, $S_{X} \propto R^{-1}$. 

\begin{figure}
	\centering\includegraphics[width=0.99\columnwidth]{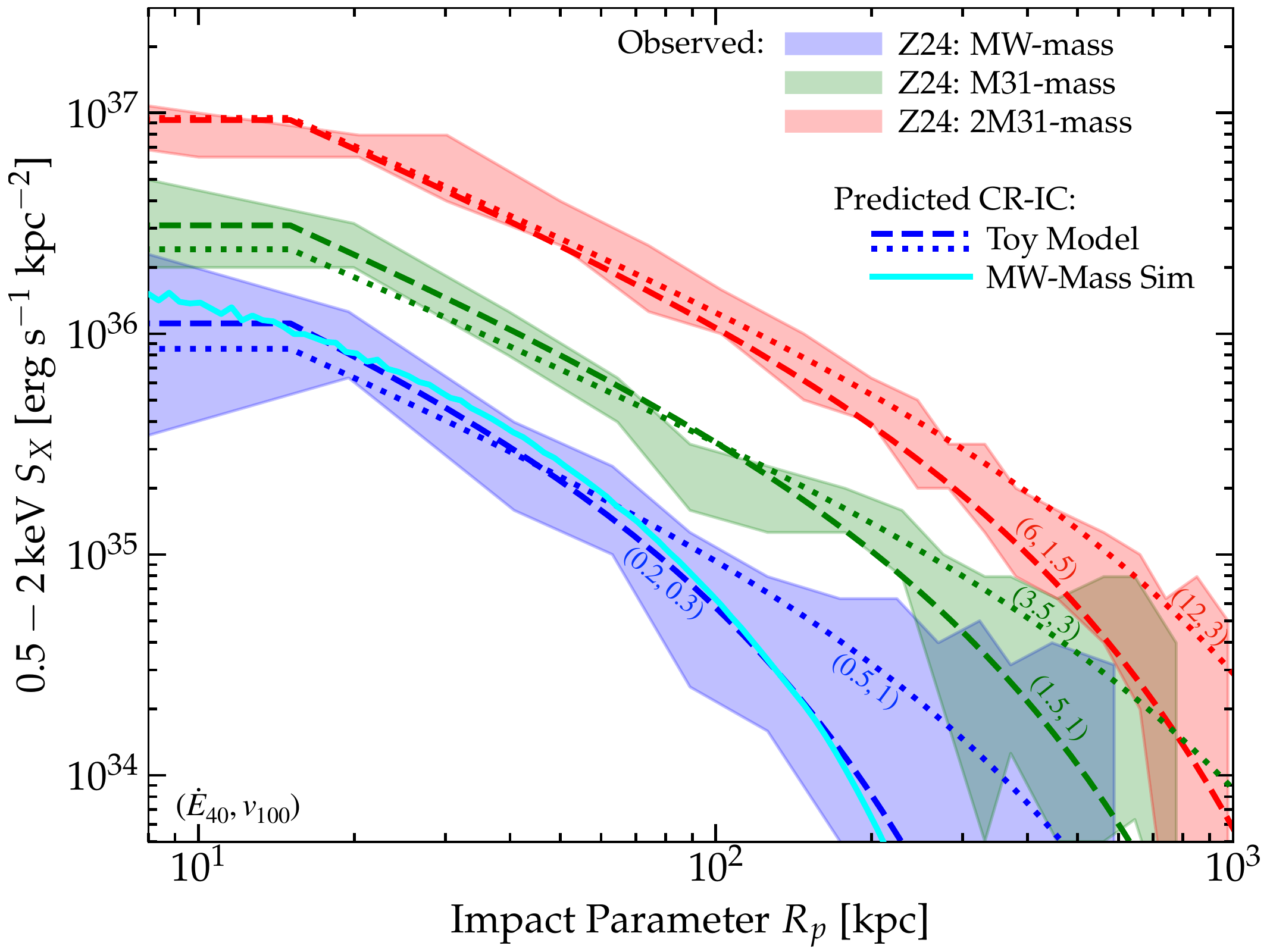} 
	\caption{Observed soft X-ray diffuse CGM emission, compared to expected IC from CRs. 
	Shaded range shows {\em eROSITA} stacked mean CGM emission at $0.5-2\,$keV, around Milky Way (MW; $10^{10.5-11}\,M_{\odot}$), Andromeda (M31; $10^{11-11.25}\,M_{\odot}$), and twice M31 (2M31; $10^{11.25-11.5}\,M_{\odot}$) stellar-mass central, isolated galaxies (after subtracting point sources/XRBs/AGN and backgrounds; \citealt{zhang:2024.hot.cgm.around.lstar.galaxies.xray.surface.brightness.profiles}). 
	Lines compare the predicted CR-IC emission, using the simple analytic model in \S~\ref{sec:basic} (with IC losses accounted for per \S~\ref{sec:losses}), for values of the injection rate parameter $\dot{E}_{\rm cr,\,\ell} = \dot{E}_{40}\,10^{40}\,{\rm erg\,s^{-1}}$ and CR streaming speed $v_{\rm st,\,eff}=v_{100}\,100\,{\rm km\,s^{-1}}$ ({\em labeled}), expected for different mass galaxies (matching colors).
	We also compare the prediction from a fiducial cosmological CR-MHD simulation ({\em cyan line}) of a MW-mass galaxy (which used a simple CR diffusion+streaming model calibrated to LISM CR data) from \citet{hopkins:cr.mhd.fire2,hopkins:2020.cr.transport.model.fx.galform}. 
	\label{fig:profiles}}
\end{figure}

This comparison is shown in Fig.~\ref{fig:profiles}, where we compare the observed $S_{X}$ around galaxies of different masses\footnote{Here and throughout, we stress that we always focus on the diffuse CGM emission observed, after subtracting the central galaxy, AGN, XRBs, and other point-source contributions as well as backgrounds. Of course there are observational uncertainties in the background subtraction and point source+PSF modeling, discussed in \citet{zhang:2024.hot.cgm.around.lstar.galaxies.xray.surface.brightness.profiles} and other papers \citep[][]{popesso:2024.stacking.on.halo.mass.extended.xray.contamination.big.issue}, but as argued in \citet{chadayammuri:2022.compare.sims.xray.cgm.profiles.vs.mass,shreeram:2025.want.to.fit.erosita.w.illustris.have.to.change.halo.masses.and.renormalize.satellites.and.assume.central.psf.not.subtracted.and.psf.wrong.and.2x.count.satellite.lum.w.central.and.assume.all.at.third.solar.metal,vladutescu:2025.magneticum.erosita.profiles.xrb.contributions} the physical contribution of these is small outside a few percent of $R_{\rm vir}$, and even assuming worst-case PSF mis-modeling they are unlikely to change the conclusions at $\gtrsim 40\,$kpc.} to the predictions of this toy model, adopting a couple different reasonable choices of the fiducial parameters $\dot{E}_{\rm cr,\,\ell}$ and $v_{\rm st,\,eff}$. For each model prediction we assume the median quoted redshift of the galaxies in that stellar mass interval from \citet{zhang:2024.hot.cgm.around.lstar.galaxies.xray.surface.brightness.profiles}, namely $\langle z \rangle \approx (0.08,\,0.12,\,0.15)$ for (MW, M31, 2M31), and calculate $f_{\rm loss}$  self-consistently following \S~\ref{sec:losses}, for the given $v_{\rm st,\,eff}$. 
We stress that these are not fits; we simply show a couple plausible values to illustrate the similarity of the expected and observed profiles.\footnote{The simple expression in Eq.~\ref{eqn:sb.oom.model} assumes point-source injection of CRs with constant $v_{\rm st,\,eff}$, and so clearly breaks down at small $R$ around the ISM of galaxies. In Fig.~\ref{fig:profiles} we simply truncate the profiles at $\sim 15\,$kpc to represent this, but note we obtain a very similar result if we adopt instead a ``streaming plus diffusion'' model with constant $\kappa_{\rm eff} \sim 1-3 \times 10^{29}\,{\rm erg\,s^{-1}\,cm^{-2}}$ dominant when $\kappa_{\rm eff} \gtrsim v_{\rm st,\,eff} \,R$, or if we model a finite uniform-density source distribution of size $\sim 10\,$kpc, as is commonly assumed in codes like GALPROP. In any case, this is only important at ISM radii, not in the CGM.}

In Fig.~\ref{fig:profiles} we also compare the ``fiducial'' {\bf m12i} simulation from \citet{hopkins:cr.mhd.fire2,hopkins:2020.cr.transport.model.fx.galform}. This is a cosmological, multi-physics, galaxy+star formation simulation following the evolution of a MW-mass galaxy at high resolution with radiative cooling, magnetohydrodynamics, star formation, and stellar feedback using the Feedback In Realistic Environments (FIRE-2; \citealt{hopkins:2013.fire,hopkins:fire2.methods}) methods. Therein, we model CRs as a single ultra-relativistic fluid obeying a simple transport model (which includes advection, streaming at the \Alf\ speed, and diffusion with a constant parallel $\kappa$ fixed to the LISM-inferred value). Since the simulation only follows the total CR energy (dominated by GeV protons), we assume an LISM-like proton-to-electron spectrum and correct for losses using the travel time $t_{\rm travel}$ estimated in-code for the GeV protons. We stress this is just one galaxy, with one (somewhat arbitrary, but empirically-motivated) CR transport model -- it is not our intention to make a rigorous comparison here. But it shows that a profile similar to that observed emerges naturally from the simplest CR transport models which have been shown to reproduce a wide variety of LISM observations in the MW \citep[see][]{chan:2018.cosmicray.fire.gammaray,chan:2021.cosmic.ray.vertical.balance,ji:fire.cr.cgm,hopkins:2020.cr.transport.model.fx.galform,hopkins:cr.multibin.mw.comparison}.

Also below we show that the predicted extended thermal emission (what has been almost exclusively invoked, including by the observers themselves, to explain the CGM emission observed in eROSITA) is much smaller than the IC emission shown (see also \citealt{vandevoort:sz.fx.hot.halos.fire}). 
This is consistent with observational upper limits to the thermal emission at $\sim 100$\,kpc implied by X-ray absorption studies \citep[e.g.][]{yao:2010.chandra.upper.limits.warm.hot.cgm.gas.in.mw.mass.halos}. These are discussed in \S~\ref{sec:thermal}, and more detailed  comparison of the IC versus thermal X-ray emission in the CGM (and comparisons with the total X-ray emission including the central galaxy and XRBs) from a larger suite of FIRE simulations is studied in \citet{lu:2025.cr.transport.models.vs.uv.xray.obs.w.cric}.

Note that for our purposes, the entire ISM/galaxy is the effective CR source region. So e.g.\ stronger losses from synchrotron or Coulomb/ionization interactions {\em within} the ISM (e.g.\ near SNe remnants), or small-scale fluctuations in CR transport physics, are effectively ``built into'' the initial spectrum we assume \citep[see discussion in][]{ponnada:2023.fire.synchrotron.profiles}. Hence adopting an LISM-like spectrum (appropriate for CRs already at the outskirts of the galaxy in the diffuse ISM, leaking into the CGM), rather than some theoretical CR acceleration spectrum at SNe shocks. This is also demonstrated quantitatively in the simulations above, calculations like \citet{zhao:2021.spatially.dependent.cr.propagation.disk.halo.models,korsmeier:2021.light.element.requires.halo.but.upper.limit.unconfined,dimauro:2023.cr.diff.constraints.updated.galprop.very.similar.our.models.but.lots.of.interp.re.selfconfinement.that.doesnt.mathematically.work,delatorre.luque:2024.gas.models.of.galaxy.key.for.scale.height.but.need.halo.for.crs}, and in Fig.~\ref{fig:profiles}: those agree very well with the observed far infrared-radio (synchrotron) correlation of galaxies (with the emission there coming from denser regions in the ISM; \citealt{ponnada:2023.fire.synchrotron.profiles,ponnada:2024.fire.fir.radio.from.crs.constraints.on.outliers.and.transport}), while also producing the extended CGM profiles shown. But we note below that our results are not particularly sensitive to that initial spectrum.

\subsubsection{Spectra and More Detailed Emissivity Calculations}
\label{sec:detailed}

Fig.~\ref{fig:spectrum} shows the resulting emissivity as a function of energy, properly integrating over the full CMB spectrum, convolved with the IC emissivity for each CR electron and positron, and then integrating over a full CR spectrum. For the CR spectrum, we show two different choices: first, we directly adopt a best-fit to the observed LISM spectrum, combining Voyager, Pamela, AMS-02, and other data sets as in \citet{cummings:2016.voyager.1.cr.spectra,2018PhRvL.120z1102A,bisschoff:2019.lism.cr.spectra}, as fit and parameterized in \citet{hopkins:cr.multibin.mw.comparison}. The positron correction is also included but is small.\footnote{There is some factor $\sim 2$ uncertainty in the exact shape of the LISM spectrum at $\sim 1$\,GeV owing to corrections for Solar modulation and lack of sensitivity of Voyager at these energies; but this is just implicit in $e_{\rm cr,\,\ell}$. Likewise for the details of the positron correction.} The second choice attempts to correct for IC losses for higher-energy CRs, as described in \S~\ref{sec:losses}.

As expected the emissivity is large near $\sim1\,$keV. For a Solar-neighorhood CR spectrum, higher-energy CRs contribute significantly and make the spectrum harder from $\sim 1-10\,$keV, with a spectral slope $dN/dt\,dE \propto E^{-\Gamma}$ with $\Gamma\sim 2.5$. But loss corrections likely make this spectrum much softer at higher energies. From this full calculation, we can revise our order-of-magnitude calculation of the emissivity in Eq.~\ref{eqn:emissivity.oom}, integrated specifically in a given band. For $0.5-2\,{\rm keV}$ (matching the eROSITA data) we obtain $\epsilon_{X}(0.5-2) \sim 2.6 \times 10^{34}\,{\rm erg\,s^{-1}\,kpc^{-3}}\,(e_{\rm cr,\,\ell}/{\rm 0.02\,eV\,cm^{-3}})\,(1+z)^{4}$ -- nearly identical to equation \ref{eqn:emissivity.oom} for $E_{\rm cr,\,GeV}\sim1$. The pre-factor in this estimate of the emissivity is no smaller than $\sim 1.5 \times 10^{34}$  (given the same {\em integrated} $e_{\rm cr,\,\ell}$) accounting for different estimates of the LISM CR spectrum. For these reasons, our conclusions are reasonably insensitive to the ``near ISM'' spectrum (LISM here), within the range implied for other MW-mass galaxies ISM from synchrotron \citep{lacki:2010.fir.radio.conspiracy,lacki:2010.fir.radio.highz.synchrotron,fletcher:RM.maps.from.synchrotron,basu:2015.synchrotron.spectral.index.and.cr.properties.nearby.sf.galaxies,beck:2015.b.field.review,beck:2019.polarization.angle.inferred.Bfield.ordering,ponnada:2023.fire.synchrotron.profiles,ponnada:2023.synch.signatures.of.cr.transport.models.fire} and $\gamma$-ray observations \citep{lacki:2011.cosmic.ray.sub.calorimetric,lopez:2018.smc.below.calorimetric.crs,zhang:2019.new.cosmic.ray.compilation.vs.calorimetry.sub.calor,tibaldo.2021:diffuse.gamma.ray.cr.profile.constraints,kronecki:2022.cosmic.ray.gamma.ray.review.not.calorimeters}: (1) these differences still imply most of the CR energy is broadly in the range $\sim 0.1-10$\,GeV (so the broad-band soft X-ray emissivity is similar for a given $e_{\rm cr,\,\ell}$), and (2) the dependence on the ``initial'' spectral slope (even experimenting with much shallower slopes like $dN/dE_{\rm cr} \propto E_{\rm cr}^{-2.2}$) is rapidly diminished as inverse Compton losses suppress the high-energy CRs.

Even though the emissivity in Figure \ref{fig:spectrum} is broadly thermal near the peak, a key difference between IC and thermal emission is that the former is of course pure continuum, while thermal emission at sub-keV temperatures is dominated by lines, even at low metallicities.    We do in fact expect some quite weak line emission in the spectrum, even if the emission is IC-dominated and the gas is actually uniformly cool ($<10^{6}\,$K) and low-density ($n < 10^{-4}\,{\rm cm^{-3}}$), so that the actual thermal contribution to the observed $\sim 0.5-2$ keV emission is negligible.  The line emission in that limit will come from a mix of fluorescence/recombination/Auger cascades from both photo-excitation by the meta-galactic X-ray background (which still dominates the observed intensity in all but the innermost halo; \citealt{zhang:2024.erosita.hot.cgm.around.lstar.galaxies.detected.and.scaling.relations}), as well as excitation and ionization by the cosmic rays themselves (not totally negligible for the CR energy densities here, as the integrated CR excitation rate for more-neutral Fe can exceed $\gtrsim 10^{-16}\,{\rm s}^{-1}\,(e_{\rm cr,\,tot}/{\rm eV\,cm^{-3}})$; \citealt{chen:1985.cr.cross.sections.iron.ionization,kaastra:1993.multiple.collisional.ionization.calcs,kovaltsov:2001.impact.ionization.cross.sections.iron}). Detailed calculations of these will depend on the temperature, metallicity, and combination of CR and X-ray background spectra and cross-sections for arbitrary electron and fluorescence yields, which are quite complicated (and for some of the species, not well-understood), but given the order-of-magnitude cross-sections in those papers, direct excitation of the salient lines by CRs should only be much larger than photo-excitation by the X-ray background for quite high CR energy densities $\gtrsim 0.01\,{\rm erg\,cm^{-3}}$, much higher than we infer from the emission at CGM radii, implying that this is a small correction to the X-ray emission in the broad-band stacks (what we care about here). 

At present, the only published detection of the diffuse CGM ($\sim 100\,$kpc) X-ray emission around MW-Andromeda mass galaxies is in broad-band ($0.5-2$\,keV), stacked soft X-ray data \citep[see][and discussion in \S~\ref{sec:thermal}]{anderson:2013.rosat.extended.cgm.xray.halos,zhang:2024.hot.cgm.around.lstar.galaxies.xray.surface.brightness.profiles}, so there is no detailed CGM spectral information to distinguish the different lines in Fig.~\ref{fig:spectrum}. We show the emissivity as a function of photon energy to highlight that the IC emission is surprisingly thermal-like (especially with IC losses accounted for), and to demonstrate that the emission peaks in the salient X-ray bands. As discussed below, future measurements of (or non-detections of) diffuse-gas line emission with X-ray microcalorimetry (e.g.\ {\em XRISM} or {\em ATHENA}) may be able to place additional useful constraints on the CR-IC scenario. Since the line emission is sensitive to the product $\sim n_{\rm gas}^{2}\,Z$, upper limits or sufficiently-weak detections would further rule out cooling/line emission as the source of the soft X-ray halos at $\gtrsim 100\,$kpc (as already implied by existing X-ray absorption measurements from \citealt{yao:2010.chandra.upper.limits.warm.hot.cgm.gas.in.mw.mass.halos,ponti:2023.erosita.supervirial.gas.close.to.galaxy.cgm.low.metal.and.low.density}, discussed further in \S~\ref{sec:thermal}). Ruling out CR-IC as the source of the emission is more challenging as it would require resolving almost all of the soft X-ray luminosity into individually detected lines, at very low surface brightness beyond $\gtrsim 100\,$kpc around a significant number of MW-mass galaxies. Even if line cooling were the only source of emission, this is challenging, as most of the cooling luminosity emerges in pseudo-continuum from overlapping lines (e.g.\ Fe L-shell emission) while these instruments are primarily sensitive to a few spectrally-isolated narrow lines \citep{kraft2023lineemissionmapperlem,zuhone2024propertieslineofsightvelocityfield}.

\subsubsection{Redshift Effects \&\ Measuring CR Energy Density}
\label{sec:redshift}

Note that there are two redshift-dependencies which appear in our IC scaling (\S~\ref{sec:basic}), both arising from the CMB: (1) the peak of the emissivity in the rest frame scaling $\propto (1+z)$, and (2) the surface brightness scaling $\propto (1+z)^{4}$. What is worth noting is that in the {\em observed} frame, these are exactly canceled by cosmological redshift and surface-brightness dimming, respectively. This means that (1) the peak of the {\em observed} X-ray emission will be at $\sim$\,keV:
\begin{align}
h \nu^{\rm obs}_{\rm peak} \sim {\rm keV}\ ,
\end{align} 
and (2) the observed/apparent surface brightness (for the same {\em physical} parameters in Eq.~\ref{eqn:sb.oom}) will be {\em independent} of the source redshift: 
\begin{align}
\frac{S_{X,\,{\rm 0.5-2\,keV}}^{\rm obs}}{\rm 10^{37}\,erg\,s^{-1}\,kpc^{-2}} &\sim f_{0} 
\left( \frac{e_{\rm cr,\,\ell}}{ 0.02\,{\rm eV\,cm^{-3}}}\right) \left(\frac{R}{100\,{\rm kpc}}\right)\ ,
\end{align}
where the range of pre-factors $f_{0} \sim 0.3-0.8$ accounts for the range of more detailed spectral shape calculations in \S~\ref{sec:detailed}. 
The soft X-ray bands are therefore always optimal for observing IC X-ray emission off of the CMB, assuming most of the CR energy resides in $\sim$GeV CRs. 

\begin{figure}
	\centering\includegraphics[width=0.99\columnwidth]{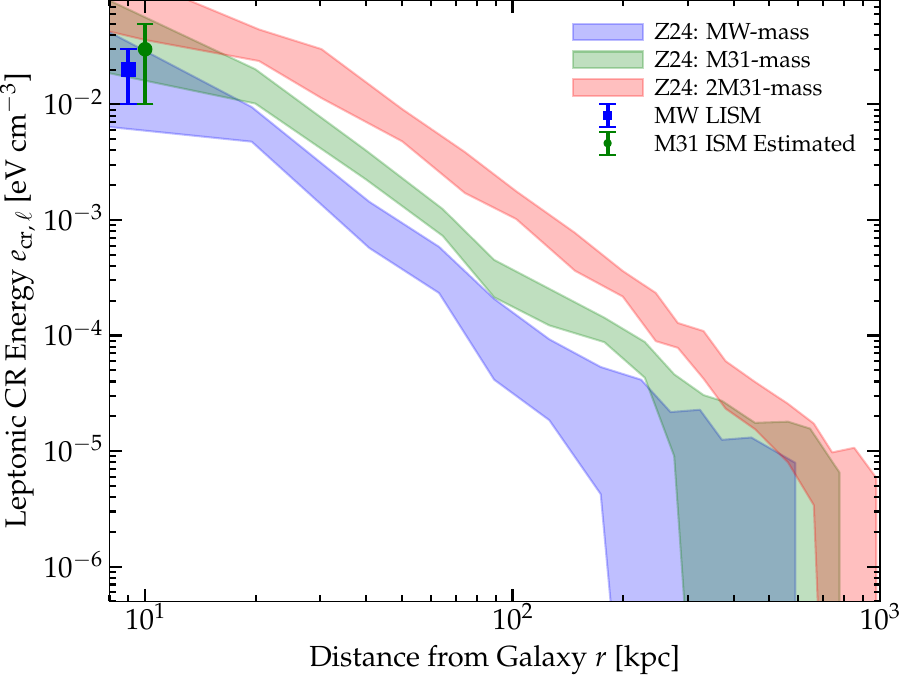} 
	\caption{Leptonic CR energy density around $\sim$\,GeV required to explain the observed soft X-ray surface brightness, for each of the three observed stacks of $S_{X}$ in Fig.~\ref{fig:profiles} ({\em shaded}; deprojecting per \S~\ref{sec:redshift}, but approximately given by Eq.~\ref{eqn:ecr}). 
	The central (galactic) values are similar to the observed LISM values \citep[][MW point with error bars]{bisschoff:2019.lism.cr.spectra} in the MW-mass systems, and those inferred from multi-wavelength modeling of M31's synchrotron plus $\gamma$-ray emission \citep[][M31 point]{lacki:2010.fir.radio.conspiracy}, with a power-law falloff and eventual cutoff from IC losses at larger radii.
	\label{fig:ecr}}
\end{figure}

If we assume IC is the dominant emission mechanism, then, measuring the CGM $\sim 1\,$keV surface brightness {\em directly} measures the CR lepton energy density around $\sim $\,GeV energies, in a unique manner unlike any other constraint (like synchrotron or $\gamma$-ray constraints, which are strictly degenerate with the magnetic field strengths and gas densities and depend qualitatively differently on redshift). In other words, one can directly read off:
\begin{align}
\label{eqn:ecr} \frac{e_{\rm cr,\,\ell}}{ 0.02\,{\rm eV\,cm^{-3}}} \sim \frac{S_{X,\,{\rm 0.5-2\,keV}}^{\rm obs}}{f_{0} 10^{37}\,{\rm erg\,s^{-1}\,kpc^{-2}}}\,\frac{100\,{\rm kpc}}{R}\ .
\end{align}
This scaling, applied to the observed $S_{X}$, is shown in Fig.~\ref{fig:ecr}.\footnote{Our $e_{\rm cr,\,\ell} \propto S_{X}/R$ scaling above assumes a roughly power-law decrease of $S_{X}$ with $R$, which technically becomes invalid at small $R$ around the ISM ($\ll 20\,$kpc) where the observed profiles become flat in Fig.~\ref{fig:profiles}. More accurately, in Fig.~\ref{fig:ecr} we de-project assuming spherical symmetry with the standard Abel integral. We obtain a nearly-identical result if we de-project assuming the best-fit spherical $\beta$ profile model in \citet{zhang:2024.hot.cgm.around.lstar.galaxies.xray.surface.brightness.profiles}. But in any case this is only important at $\ll 20\,$kpc.}
In theoretical {\em models} for the CR propagation, this in turns constrains the product $f_{\rm loss} \dot{E}_{\rm cr,\,\ell}/v_{\rm st,\,eff}$ (Eq.~\ref{eqn:sb.oom.model}), $f_{\rm loss}\,\dot{E}_{40}/v_{100} \sim (S_{X,\,{\rm 0.5-2\,keV}}^{\rm obs} / 10^{35.3}\,{\rm erg\,s^{-1}\,kpc^{-2}})\,R_{100}$. 

Note the rest-frame scaling of $S_{X}$ with $(1+z)^{4}$ means the IC loss time also scales $\propto (1+z)^{-4}$, which means (\S~\ref{sec:losses}) the maximum extent of the IC halo (before being loss-limited) scales will be smaller at higher redshift (modulo possible scaling of varying streaming/diffusion speeds and reacceleration efficiencies, but these should not evolve so rapidly in most models\footnote{For example, halo virial and ``backsplash'' radii scale as $\propto (1+z)^{-1}$, central densities and scale radii evolve more weakly \citep{bullock:concentrations}, and virial temperatures scale $\propto (1+z)^{1/2}$ \citep{bryan.norman:1998.mvir.definition}, at fixed $M_{\rm vir}$.}). So while the halo {\em luminosity} remains independent of redshift, the effective maximum extent (in physical units) will be smaller at high redshifts -- presenting a potential test of the IC scenario. This redshift-dependence will be quite difficult to detect, however, for two reasons. First, it would require exceptional angular resolution: noting that {eROSITA} only barely detects the MW-mass halos above their PSF wings at $z\sim 0.08$ \citep{zhang:2024.hot.cgm.around.lstar.galaxies.xray.surface.brightness.profiles}, the combination of shrinking physical size and rapidly-rising angular diameter distance means that measuring this predicted trend at $z \gtrsim 0.5$ would require more than an order-of-magnitude better angular resolution (e.g.\ a $\lesssim 1''$ PSF). Second, the X-ray background also scales $\propto (1+z)^{4}$, or equivalently surface-brightness dimming scales as $(1+z)^{-4}$; so if one had a halo powered by thermal emission at higher redshift, its {\em detectable} size above some threshold relative to the backgrounds would also shrink. To properly compare, one would need signal-to-noise scaling as $\propto (1+z)^{4}$ at higher redshifts. These requirements push the limits even of potential future missions like AXIS.

\subsection{(Inverse Compton) Losses}
\label{sec:losses}

The IC lepton loss timescale $t_{\rm loss} \equiv p_{\rm cr}/\dot{p}_{\rm cr}^{\rm IC} \sim 1.2\,(1+z)^{-4}\,E_{\rm cr,\,GeV}^{-1}$\,Gyr (for the CRs with energies actually of interest to produce CR-IC in the observed bands) should eventually become comparable to or shorter than the CR escape/diffusion/travel time to some radius $t_{\rm travel} \equiv R/v_{\rm st,\,eff} \sim  1\,(R/100\,{\rm kpc})\,(100\,{\rm km\,s^{-1}}/v_{\rm st,\,eff})$\,Gyr, i.e.\ $\tau_{\rm loss} \equiv t_{\rm travel}/t_{\rm loss} \sim E_{\rm cr,\,GeV}\,R_{100}\,(1+z)^{4}/v_{100}$. This depends on the highly-uncertain CGM streaming/diffusion speeds, so for faster CR transport it could be a small correction, but generically we would expect that at sufficiently large radii and CR energies, $\tau_{\rm loss} \gg 1$, and the CR spectrum will be suppressed.\footnote{Note the exact form of the suppression is modified if the transport is diffusive, as compared to streaming-like (\citealt{quataert.hopkins:2025.crs.massive.halos.blowout.rvir.cosmology.constraints}), but if the diffusivity increases with radius or if there is super-diffusion, it becomes more like the streaming case again (\citealt{liang:2025.leaky.boxes.levy.flights.modeling.crs.transport}, Ponnada et al., in prep.). More general cases can be calculated numerically from the gyro-averaged CR equations \citep{hopkins:m1.cr.closure,thomas:2021.compare.cr.closures.from.prev.papers}, but for the simple case here are variations of solutions to a Sturm-Liouville-type equation (\S~\ref{sec:basic}), and for the same median travel time, these only influence the details of precisely ``how steep'' the suppression is, once it occurs.}

It is straightforward to verify from standard scalings \citep{1965AnAp...28..171G,blumenthal:1970.cr.loss.processes.leptons.dilute.gases,1972Phy....60..145G,ginzburg:1979.book,rybicki.lightman:1979.book} that -- for the energies of interest -- other CR losses in the CGM/IGM are negligible, including catastrophic and pionic (insignificant for leptons at $\sim$\,GeV), synchrotron (see \S~\ref{sec:radio.gamma}), bremsstrahlung ($t_{\rm loss}^{\rm brem} \sim 3000\,{\rm Gyr}/n_{-5}$), Coulomb ($t^{\rm Coul}_{\rm loss} \sim 5000\,{\rm Gyr}\,E_{\rm cr,\,GeV}/n_{-5}$), and ionization ($t_{\rm loss}^{\rm ion} \sim t^{\rm Coul}_{\rm loss}/f_{\rm neutral} \gg t^{\rm Coul}_{\rm loss}$).  
The ``streaming losses'' ($\sim {\bf v}_{A} \cdot \nabla P_{\rm cr}$; \citealt{wentzel.1969.streaming.instability,kulsrud.1969:streaming.instability}) and ``adiabatic/$PdV$ work'' ($\sim P_{\rm cr}\,\nabla \cdot {\bf u}_{\rm gas}$; \citealt{jokipii:1966.cr.propagation.random.bfield,Voelk1973}) terms, which generalize the classic ``turbulent reacceleration terms'' ($\lesssim (v_{A}^{2}/3\,\kappa_{\rm eff})\,P_{\rm cr}$; references above) can be non-zero but in numerical simulations which explicitly model them, they are found on average to rarely be more than $\mathcal{O}(1)$ corrections to the CR energy density and emissivity, depending on the structure of the turbulence and magnetic fields in the outer halo \citep{chan:2018.cosmicray.fire.gammaray,buck:2020.cosmic.ray.low.coeff.high.Egamma,thomas:2022.self-confinement.non.eqm.dynamics,hopkins:cr.multibin.mw.comparison}. But they could modify the details of the outer cutoff and should be studied in future work.

Note in terms of heating the {\em gas}, the IC radiation is rather inefficient, though a small fraction of the lowest-energy IC photons could go into ionizing higher metal-line transitions. Depending on the details of the CR-gas interactions at the gyro-scale, the CR ``streaming heating'' term will likely be the most important for heating the gas (references above)\footnote{This is because the volumetric heating rate from streaming losses scales as $\sim |v_{A} \nabla P_{\rm cr}|$ \citep{1975MNRAS.173..245S}, independent of gas density, while other processes like Coulomb, ionization, bremsstrahlung, and heating from pion production by CR protons and other hadrons all scale $\propto e_{\rm cr} n_{\rm gas}$, in the diffuse outer CGM where $n_{\rm gas}$ is low (giving rise to the long loss timescales above).}, though simulations with similar $e_{\rm cr}$ as implied by the X-ray halos generally find this is sub-dominant to gas cooling by factors of $\sim 10-100$ in low-mass ($\lesssim 10^{13}\,M_{\odot}$) halos \citep{ji:20.virial.shocks.suppressed.cr.dominated.halos,ji:2021.cr.mhd.pic.dust.sims,martin.alvarez:2023.mhd.cr.sims.synch.maps.similar.emission.regions.conclusions.to.fire.ponnada.papers.but.very.different.methods,dacunha:2024.synthetic.synchrotron.problems.with.obs.interp}.

\subsubsection{Modification of the CR Spectrum}
\label{sec:ic.spectrum}

The evolution of a CR lepton with initial energy $E_{\rm cr,\,0}$ and IC loss time $t_{\rm loss,\,0}$ in a uniform radiation background with only IC losses has the simple solution: $E_{\rm cr} = E_{\rm cr,\,0}/(1+t/t_{\rm loss,\,0})$. If we assume a population of CRs is injected at time $t=0$, and neglect other losses, then the CR spectrum evolves in time as  $dn_{\rm cr}[E_{\rm cr}]/d\ln{E_{\rm cr}} = dn_{\rm cr}[E^{0}_{\rm cr}]/d\ln{E^{0}_{\rm cr}}\, (1 + t/t_{\rm loss}[E_{\rm cr}^{0}])$, which depends only on the parameter $\tau \equiv t/t_{\rm loss,\,0}$. The IC-modified spectra are shown in Fig.~\ref{fig:spectrum} for different values of the parameter: 
\begin{align}
\label{eqn:crit.loss} \tau_{\rm loss}^{0} \equiv \frac{t_{\rm travel}}{t_{\rm loss}}{\Bigr|}_{\rm GeV} \sim \frac{0.8}{(1+f_{\rm reacc})}\, \frac{R_{100}}{v_{100}}\,(1+z)^{4}
\end{align}
Note the spectrum at a given radius in the CGM formally depends (1) on how we assume $v_{\rm st}$ depends on CR energy;\footnote{We have compared assuming a constant streaming speed, or constant diffusivity, or allowing $v_{\rm st}$ to depend on energy with some dependence (faster diffusive transport of high-energy CRs) like that inferred in the LISM, $v_{\rm st} \propto 1 + E_{\rm cr,\,GeV}^{1/2}$, and find these make only modest quantitative differences to the spectra at the radii of interest, and do not change any of our qualitative results.}  (2) the mix of streaming and diffusion and inhomogenous CR transport, which will broaden the ``cutoff'' in the IC-modified spectrum because of different travel times at a given radius; and (3) the mix of streaming (CR gyro-resonant instability) losses, adiabatic CR gains/losses, and turbulent and diffusive re-acceleration (parameterized above with $f_{\rm reacc}$) during CR transport  \citep[see e.g.][]{gaggero:2015.cr.diffusion.coefficient,hopkins:cr.transport.constraints.from.galaxies,hopkins:2021.sc.et.models.incompatible.obs,korsmeier:2022.cr.fitting.update.ams02,dimauro:2023.cr.diff.constraints.updated.galprop.very.similar.our.models.but.lots.of.interp.re.selfconfinement.that.doesnt.mathematically.work}. All of these effects generally introduce only $\mathcal{O}(1)$ changes at most in the total emissivity at $\sim 1\,$keV, since they mostly just modify how rapidly the already-rapid cutoff in the spectrum occurs. 
Regardless of how these details modify our calculation, we obtain the expected result that when $\tau_{\rm loss}^{0} \gg 1$, the CR spectrum must be strongly modified from its injection spectrum and most of the energy has been lost (so the emissivity at larger $R$ must be small). 

Note, as discussed above, that as IC losses modify the spectrum, it becomes softer and more similar to thermal spectra. The measured emissivity spectrum for loss parameter $\tau_{\rm loss}^{0} = 1$, for example, would look similar to gas with $T \approx 1.5\times 10^{7}\,$K, while at $\tau_{\rm loss}^{0} = 10$, it closely resembles the spectrum of $\sim 10^{6}\,$K gas. At present, observations of the diffuse emission in MW-Andromeda mass galaxies cannot constrain the spectra (just the integrated soft X-ray surface brightness), but in principle this is relevant for future searches, and it informs our predictions for associated harder X-ray and synchrotron emission below (\S~\ref{sec:radio.gamma}).

\begin{figure}
	\centering\includegraphics[width=0.99\columnwidth]{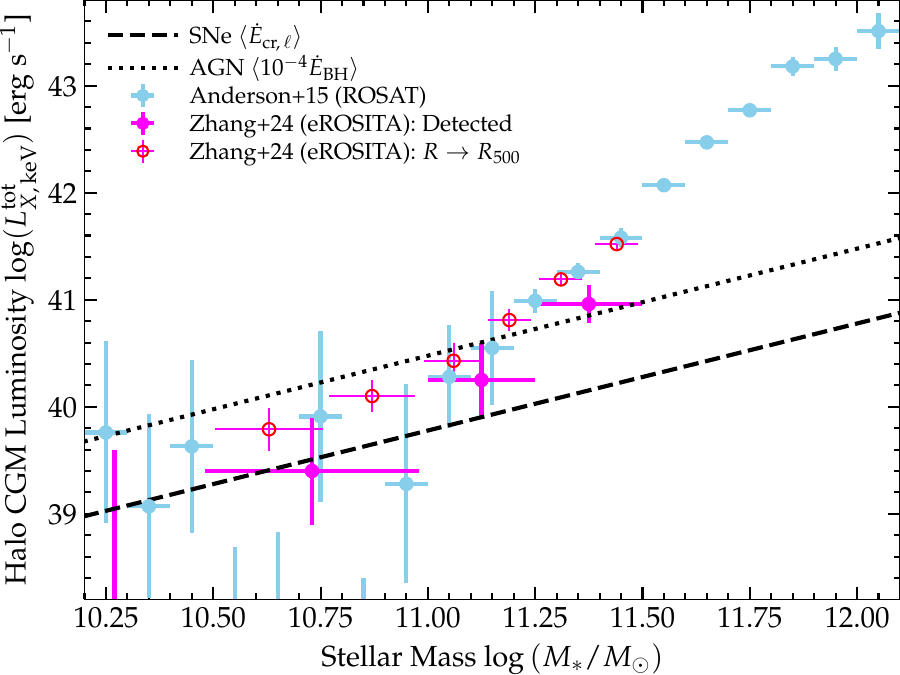} 
	\caption{Total CGM halo soft X-ray luminosity, $L_{\rm X,\,kev}^{\rm tot} = \int S_{X}\,dA$, as a function of central galaxy stellar mass $M_{\ast}$, compared to the expected IC luminosities. Observed values from are shown from stacked ROSAT data (\citealt{anderson:2015.rosat.xray.halo.stacking.scaling.relations} CGM only-emission after subtracting background+galaxy+point-source/XRB contributions), as well as eROSITA, using the same brightness profiles from \citet{zhang:2024.hot.cgm.around.lstar.galaxies.xray.surface.brightness.profiles} in Fig.~\ref{fig:profiles} and integrating out to radii where any excess emission is detected (i.e.\ allowing for an IC cutoff; {\em filled}; ``detected''), or using the stacking method in \citet{zhang:2024.erosita.hot.cgm.around.lstar.galaxies.detected.and.scaling.relations} which extrapolates the inferred profiles from smaller radii out to $R_{500}$ in all cases ({\em open}; ``$R\rightarrow R_{500}$''). We compare the predicted time-and-population-averaged SNe $\sim$\,GeV  leptonic CR injection rate and AGN injection rate (assuming empirical SFR-$M_{\ast}$ and $\langle M_{\rm BH} \rangle - M_{\ast}$ relations, and that $\sim 10^{-4}\,\dot{M}_{\rm BH}\,c^{2}$ goes into leptons, for reference; see \S~\ref{sec:obs}) from just the central galaxies. These scalings have factor $\sim 2-3$ systematic uncertainties. IC from standard sources (SNe and/or AGN even with very small efficiencies) can naturally explain the halo luminosities at $M_{\ast} \lesssim (1-2)\times 10^{11}\,M_{\odot}$.
	\label{fig:lum}}
\end{figure}

\subsubsection{Total Luminosity and Extent of Halos: Estimating the CR Injection Rate and Streaming Speed}
\label{sec:ltot}

Given IC losses per \S~\ref{sec:ic.spectrum}, at some radius the CR surface brightness should rapidly truncate, approximately as $\propto \exp{(-\tau_{\rm loss}^{0})}$. The more robust way of saying this is that once the integrated IC luminosity out to some radius $R$, $L_{\rm keV}(R) \equiv \int_{0}^{R} S_{X}(R)\,\pi R\,dR$, exceeds the leptonic GeV CR source injection rate $\dot{E}_{\rm cr,\,\ell}$ (modulo some order-unity correction for other losses and adiabatic/diffusive/turbulent reacceleration, $\sim 1+f_{\rm reacc}$), then the CRs must be depleted to lower energies and their IC from the CMB must shift out of the keV band,\footnote{There could be some X-ray radiation from lower-energy CRs scattering the cosmic infrared/optical backgrounds, but the photon number density of these backgrounds and corresponding X-ray emissivity is lower than for the CMB by factors of $\sim 10^{3}-10^{6}$, so we neglect them throughout.} so $S_{X}$ drops rapidly and $L_{\rm keV}$ cannot rise further. The radius where this will occur is just the radius where $\tau_{\rm loss}^{0} \gtrsim $\,a couple, i.e.\ $R \gtrsim R_{\rm crit} \sim 100\,{\rm kpc}\,(1+z)^{-4}\,v_{100}\,(1+f_{\rm reacc})/0.8$ (Eq.~\ref{eqn:crit.loss}). Another way of saying the above is that if we insert $R_{\rm crit}$ into Eq.~\ref{eqn:sb.oom.model}, then we trivially recover $L_{\rm keV}(R\rightarrow R_{\rm crit}) \sim \pi\,R_{\rm crit}^{2} S_{X}(R_{\rm crit}) \sim \dot{E}_{\rm cr,\,\ell}$. 

Therefore, if one can {\em measure} a truncation or ``cutoff'' radius where $S_{X}$ exhibits some curvature, one can directly infer the CR streaming speed (since the lifetime is known for IC against the CMB) as $v_{100} \sim R_{\rm crit}\,(1+z)^{4}\,(0.8/\tau_{\rm loss}^{0,\,\rm crit}\,(1+f_{\rm reacc}))$. This then, with Eq.~\ref{eqn:sb.oom.model}, allows one to measure $\dot{E}_{\rm cr,\,\ell}$.

Or equivalently, if one can just measure the asymptotic IC {\em luminosity} of the CGM halos, $L_{\rm keV}$, then one obtains the approximate leptonic injection rate into the CGM
\begin{align}
(1+f_{\rm reacc}) \,\dot{E}_{\rm cr,\,\ell} \sim L_{X,\,{\rm keV}}^{\rm tot} \ .
\end{align}
This is shown, as a function of galaxy mass, in Fig.~\ref{fig:lum}. 
Importantly, the observed luminosities are the CGM luminosities {\em after} subtraction of XRBs, AGN, and any other central source (e.g.\ hot gas in the galactic ISM). As shown in many studies including those plotted \citep[e.g.][]{anderson:2015.rosat.xray.halo.stacking.scaling.relations,zhang:2024.hot.cgm.around.lstar.galaxies.xray.surface.brightness.profiles}, in low-mass galaxies the total X-ray luminosity {\em including} the galaxy and all sources is dominated by a combination of HMXBs+AGN in the central $\lesssim 5$\,kpc (and simulation comparisons with these are discussed in e.g.\ \citealt{vandevoort:sz.fx.hot.halos.fire,chan:2021.cosmic.ray.vertical.balance} and \citealt{lu:2025.cr.transport.models.vs.uv.xray.obs.w.cric}).

With this estimate of $\dot{E}_{\rm cr,\,\ell}$, then from the measured $S_{X}$ {\em interior} to the outer/asymptotic radii ($R \ll R_{\rm crit}$, where the loss corrections should not be very large) we can roughly infer the CR effective streaming speed or diffusivity:
\begin{align}
\label{eqn:vstream.measured} v_{\rm st,\,eff}&(R < R_{\rm crit}) \sim \frac{\kappa_{\rm eff}(R)}{R} \sim 200\,{\rm km\,s^{-1}}\,\frac{ L_{X,\,{\rm keV},40}^{\rm tot}}{S_{X,\,35} R_{100}}\ , \\ 
\label{eqn:kappa.measured} {\kappa_{\rm eff}} &\sim R\,v_{\rm st,\,eff}(R) \sim 6 \times 10^{30}\,{\rm cm^{2}\,s^{-1}}\,\frac{ L_{\rm X,\,{\rm keV},40}^{\rm tot}}{S_{X,\,35}}\ , 
\end{align}
where $S_{X,\,35} \equiv S_{X,\,0.5-2\,{\rm keV}}^{\rm obs} / 10^{35}\,{\rm erg\,s^{-1}\,kpc^{-2}}$, $L_{X,\,{\rm keV},\,40}^{\rm tot} \equiv L_{X,\,{\rm keV}}^{\rm tot}/10^{40}\,{\rm erg\,s^{-1}}$, $R_{100} \equiv R/100\,{\rm kpc}$. 
These are shown in Fig.~\ref{fig:kappa}. The shaded range there includes the uncertainty in $S_{X}$ and $L_{X}$ (added in quadrature), but we caution it does not include potential systematic errors from simplifying assumptions (like spherical averaging).

Note that for more massive groups and clusters, we should properly account for injection from many galaxies throughout the halo, as the stellar mass and total AGN luminosity becomes increasingly dominated by the sum of galaxies rather than the BCG in clusters (and these, being injected at finite $R$, will suffer far less IC losses). But in the lower-mass halos of interest here like in the Local Group, the total satellite luminosity/mass is much lower than that of the central \citep{geha:2024.saga.dwarf.galaxies.quenched.fractions}.

\subsection{Observational Constraints: CR Energies and Streaming Speeds}
\label{sec:obs}

\begin{figure}
	\centering\includegraphics[width=0.99\columnwidth]{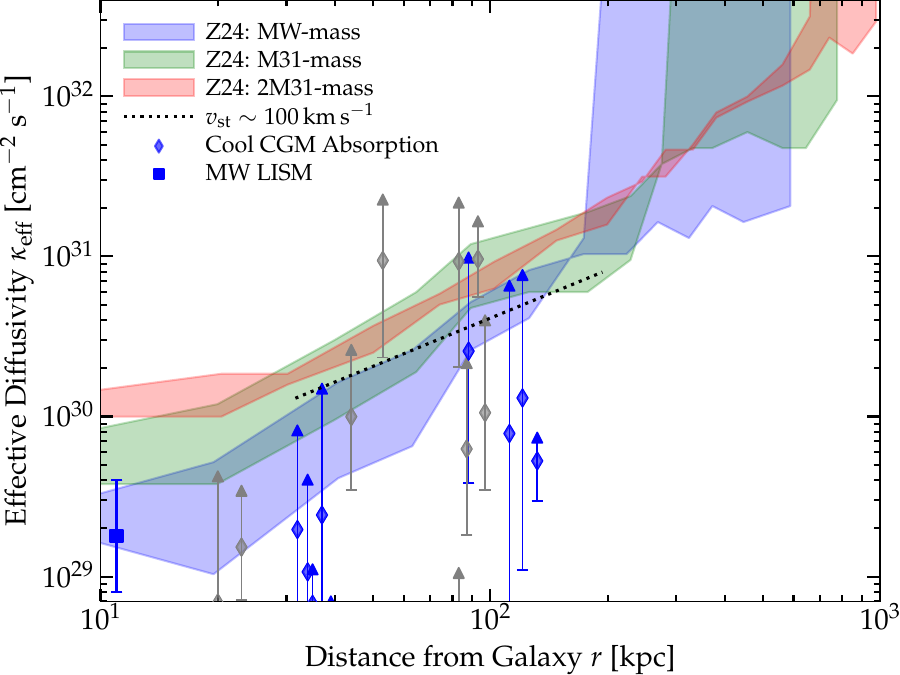} 
	\caption{Implied effective CR diffusivity $\kappa_{\rm eff}$ to explain each observed $S_{X}$ profile ({\em shaded}), given the known IC loss timescale and observed injection rate $\dot{E}_{\rm cr} \sim L_{X,\,{\rm keV}}^{\rm tot}$ (\S~\ref{sec:ltot} \&\ \ref{sec:obs}, Eq.~\ref{eqn:kappa.measured}). 
	Approaching the ISM, the values in MW/M31 are very similar to those inferred from standard Galactic CR modeling of MW LISM observations from e.g.\ Voyager, AMS-02, etc. (MW LISM point, median and range from the model sets in \citealt{delaTorre:2021.dragon2.methods.new.model.comparison,korsmeier:2022.cr.fitting.update.ams02} using GALPROP and DRAGON2). 
	At larger radii $\kappa_{\rm eff}$ rises, as generally expected in the CGM.
	We compare independent observational constraints on $\kappa_{\rm eff}$ for $\sim$\,GeV protons from a completely independent UV absorption-line method applied to cool neutral HI gas (which should not appear in the X-rays) in \citet{butsky:2022.cr.kappa.lower.limits.cgm}, colored by stellar mass ({\em blue} matching Z24's MW-mass sample, {\em grey} for masses outside this). 
	\label{fig:kappa}}
\end{figure}

\citet{zhang:2024.hot.cgm.around.lstar.galaxies.xray.surface.brightness.profiles} measure\footnote{We compare to the eROSITA data from \citet{zhang:2024.hot.cgm.around.lstar.galaxies.xray.surface.brightness.profiles} as the improved sensitivity allows much more detailed model constraints compared to older ROSAT data (which are unable to spatially-resolve the halos around lower-mass systems) in \citep{anderson:2013.rosat.extended.cgm.xray.halos,anderson:2015.rosat.xray.halo.stacking.scaling.relations}, but note that the integrated CGM/halo luminosities as a function of stellar mass (and halo sizes, where ROSAT can detect and spatially-resolve them) are broadly consistent with one another. We also focus on the stellar mass-selected samples in \citet{zhang:2024.hot.cgm.around.lstar.galaxies.xray.surface.brightness.profiles}, rather than their halo-mass selected samples, as selection in the latter is model-dependent and can give very different results depending on the scatter in the $M_{\ast}-M_{\rm halo}$ relation \citep[see e.g.][]{popesso:2024.erosita.stacking.lx.mhalo.by.halo.mass,popesso:2024.stacking.on.halo.mass.extended.xray.contamination.big.issue}, and exclusively compare their background and point-source subtracted best estimate of the diffuse halo emission. When comparing $L_{X}$ and $M_{\ast}$, we correct the stellar masses of \citet{anderson:2015.rosat.xray.halo.stacking.scaling.relations} by the same small ($0.1$\,dex) factor as \citet{zhang:2024.erosita.hot.cgm.around.lstar.galaxies.detected.and.scaling.relations} to align their definitions.} the stacked $0.5-2\,$keV CGM surface brightness around three stellar mass-selected samples of galaxies (compared in Fig.~\ref{fig:profiles}): their ``Milky Way mass'' (MW) sample ($M_{\ast} \sim 10^{10.5-11}\,M_{\odot}$), ``Andromeda mass'' (M31) sample ($M_{\ast} \sim 10^{11-11.25}\,M_{\odot}$), and ``twice Andromeda mass'' (2M31) sample ($M_{\ast} \sim 10^{11.25-11.5}\,M_{\odot}$). The median central-galaxy stellar mass in these (MW, M31, 2M31) samples is $\sim (0.55,\,1.3,\,2.2)\times 10^{11}\,M_{\odot}$. 

From the observations, we can estimate $L_{X,\,{\rm keV}}^{\rm tot}$ in Fig.~\ref{fig:lum} by integrating along the lower and upper limits on the stacked profiles out to the radii where they become undetectable against the X-ray background (where the lower limits fall below the plot). In principle a lower-level background halo could persist, but this seems to be indicative of some steepening in the profiles as shown in Fig.~\ref{fig:profiles}. We obtain $L_{X,\,{\rm keV},\,40}^{\rm tot} \sim (0.08-0.8,\, 0.8-4,\, 6-14)$ for (MW, M31, 2M31).\footnote{This is different from the method of estimating $L_{X,\,{\rm CGM}}$ in \citet{zhang:2024.erosita.hot.cgm.around.lstar.galaxies.detected.and.scaling.relations}, who extrapolate a fitted or upper limit to $S_{X}$ out to $R_{500}$, which leads to factor of a few larger $L_{X,\,{\rm CGM}}$ for the low-mass halos. It is more akin to the estimates in \citet{anderson:2015.rosat.xray.halo.stacking.scaling.relations}.} For the streaming speeds at $R \lesssim R_{\rm crit}$, we obtain at $R \sim 10\,$kpc, $v_{\rm st,\,eff} \sim (25-80,\, 80-170,\, 100-400 )\,{\rm km\,s^{-1}}$, or $\kappa_{\rm eff} \sim (0.8-2.4,\, 2-5,\, 3-12) \times 10^{29}\,{\rm cm^{2}\,s^{-1}}$ for (MW, M31, 2M31). At $R \sim 100\,$kpc, this implies $v_{\rm st,\,eff} \sim (80-160,\, 100-200,\, 120-160 )\,{\rm km\,s^{-1}}$ or $\kappa_{\rm eff} \sim (2-5,\, 3-6,\, 3.5-5) \times 10^{30}\,{\rm cm^{2}\,s^{-1}}$, shown in Fig.~\ref{fig:kappa}. Even if we allow for much larger potential systematic uncertainties, this is extremely powerful, given the lack of extragalactic constraints on CR properties in the CGM at $\gtrsim 100\,$kpc from MW-mass galaxies.

\subsubsection{Comparison to Other Empirical Constraints and Models}

The CR electron injection rates required to explain the X-ray halos of MW-Andromeda  mass galaxies via IC emission are quite similar to what is observed and expected in local galaxies from standard sources of CRs. Of course, detailed models of individual galaxies like the simulations in Fig.~\ref{fig:profiles} imply this, but it is helpful to make a simple order-of-magnitude estimate of the CRs accelerated by (1) SNe and (2) AGN (and/or associated jets/bubbles/winds), to compare the population scalings in Fig.~\ref{fig:lum}.

For SNe (1), we can sum core-collapse and prompt Ia rates (proportional to galactic star formation rates, $\dot{N}_{\rm prompt} \sim 0.014\,(\dot{M}_{\ast}/M_{\odot})$; \citealt{sukhbold:yields.and.explosion.props.dense.grid,hopkins:fire3.methods}) and delayed Ia rates (proportional to stellar mass $\dot{N}_{\rm late} \sim 1.5 \times 10^{-13}\,{\rm yr}^{-1}\,(M_{\ast}/M_{\odot})$; \citealt{maoz:Ia.rate,gandhi:2022.metallicity.dependent.Ia.rates.statistics.fire}) using the fact that for  stacked samples in these mass ranges, the {\em mean} (what matters for the stacked result) SFRs and stellar masses are well-known. For stellar mass, $\langle M_{\ast} \rangle \sim (0.55,\,1.3,\,2.2)\times 10^{11}\,M_{\odot}$ is defined by the (MW, M31, 2M31) samples in the observations; the mean SFRs are given by the ``main sequence'' ($\dot{M}_{\ast}-M_{\ast}$ correlation) average (dominated by the rapidly star-forming population) $\langle \dot{M}_{\ast} \rangle \sim (3,\,7,\,11)\,{\rm M_{\odot}\,yr^{-1}}$ \citep{pearson:2018.main.sequence.sfr.z,cooke:2023.sfr.main.sequence.evol}. If we assume the standard empirically-estimated $\sim 10^{50}\,{\rm erg}$ of total CR energy accelerated per SNe ($\sim 10\%$ of the ejecta energy), with $\sim 2\%$ of that in $\sim$\,GeV leptons, we obtain $\langle \dot{E}_{\rm SNe,\,\ell} \rangle \sim (0.3,\,0.8,\,1.2) \times 10^{40}\,{\rm erg\,s^{-1}}$. So for the lower-mass (MW, M31) samples (Fig.~\ref{fig:lum}), the observed halo luminosity is entirely consistent with the expectation from SNe-accelerated leptons. 

The AGN contribution (2) is more uncertain, but the mean bolometric AGN luminosity and accretion rates are empirically known for galaxies with these stellar masses (and/or SFRs). Here it is more important that we account for the mean, not median, since (given the large range of $L_{\rm AGN}$ at a given $M_{\ast}$) more luminous systems should dominate the injection. From \citet{torbaniuk:2021.bhar.stellar.mass.sfr.corrs,torbaniuk:2024.bhar.sfr.mstar.relations}, for these mass ranges, we have $\langle \dot{M}_{\rm BH} \rangle \sim (2.5-3,\,5.3-7.1,\,8.8-16)\times 0.001\,{\rm M_{\odot}\,yr^{-1}}$, or an available accretion energy $\langle \dot{E}_{\rm BH} \rangle \sim (1.4-1.7,\,3-4,\,5-9) \times 10^{44}\,{\rm erg\,s^{-1}}$. So powering the observed halos via AGN injection of CRs would require only a small fraction $\epsilon_{\rm cr,\,\ell} \sim (0.4-6,\,2-12,\,6-28) \times 10^{-5}$ of the accretion energy ultimately goes into accelerating leptons -- a completely plausible estimate from previous constraints in radio AGN \citep{falcke04:radio.vs.mdot,birzan:2004.radio.cavity.jet.power.vs.xray.power.correlation,allen:jet.bondi.power,mcnamara:2007.agn.cooling.flow.review.cavity.jet.power.vs.xray.luminosity.scalings.emph.compilation}. In both of these, we could also include the contribution of satellite galaxies, which should grow more important at large radii and more massive halos, but this is expected to be small for MW-M31 mass halos. 

Briefly, it is worth noting that the {\em galaxy} (not diffuse CGM) X-ray luminosity of star-forming (low-mass) galaxies observationally scales as \citep{Mineo2014}
\begin{align}
L^{\rm galaxy}_{X}(0.5-8\,{\rm keV}) &\approx 4 \times 10^{39} \, {\rm erg \, s^{-1}} \left(\frac{\dot M_{\ast}}{M_\odot \, {\rm yr^{-1}}}\right) \ , 
\label{eq:LXSFR}
\end{align}
which is expected from the rate of both SNe (creating hot gas) and XRBs (with e.g.\ \citealt{Mineo2012} attributing $\sim 2/3$ of this to resolved HMXBs). Ignoring AGN, the scalings above for CR injection give $L_{X,\,{\rm IC}} \sim \dot{E}_{\rm cr,\,\ell} \sim 10^{39}\,{\rm erg\,s^{-1}}\,(\dot M_{\ast} / {\rm M_{\odot}\,yr^{-1}})$. So the expectation is that the extended IC X-ray halo luminosity should scale with (and be within a factor of a few of) the central galaxy/XRB luminosity in star-forming galaxies, although the IC emission is much more diffuse and extended, as indeed appears to be observed.

Regarding AGN, note that we do not expect CR halos powered by BHs/AGN to necessarily correlate with {\em ongoing} AGN activity. Not only is the stack population-averaged, but the CR travel time to the radii of interest is $\sim$\,Gyr as noted above, much longer than the lifetime of AGN accretion episodes \citep{martini04,hopkins:lifetimes.letter,hopkins:lifetimes.methods,hopkins:lifetimes.interp,hopkins:lifetimes.obscuration,kelly:2010.bhmf}, so the effective CR injection luminosity is averaged over the whole duty cycle of AGN at low redshifts. Interestingly, \citet{zhang:2025.erosita.extended.halos.luminosity.versus.mass.versus.sf.quenched} recently argued that the eROSITA CGM emission of quiescent MW/M31-mass systems is either comparable to or perhaps even somewhat larger than that of star-forming galaxies (depending on how they account for the well-known fact that quiescent galaxies of similar optical luminosity reside in more massive halos; e.g.\citealt{behroozi:2019.sham.update,tinker:2021.sdss.group.finder.color.dependence}). This could be consistent with the efficiencies $\epsilon_{\rm cr,\,\ell}$ shown in Fig.~\ref{fig:lum}, which predict somewhat higher BH injection than SNe injection, given the fact that quiescent galaxies are observed to host significantly larger BH masses at the same stellar mass \citep{bluck:2014.bulge.mass.best.predictor.of.quenching,bluck:2023.quenching.sims.obs.machine.learning.key.is.bh.mass.vs.galaxy.mass.total,reines:2015.dwarf.gal.mbh.mgal.norm.dift.and.huge.scatter,terrazas:2017.quenching.correlated.specific.BH.mass,chen:2020.overmassive.bh.quenching.model,wang:2024.strong.relation.bh.mass.gas.mass.galaxies}. But the significance of the offset is unclear (compare e.g.\ \citealt{chadayammuri:2022.compare.sims.xray.cgm.profiles.vs.mass}), and more detailed forward-modeling is needed since this should depend, in detail, on when those BHs were active.

For the implied streaming speeds/diffusivities, our estimator is approximate of course, but the physical uncertainties in theoretical predictions are much larger \citep{zweibel:cr.feedback.review,hopkins:cr.transport.constraints.from.galaxies,hopkins:2021.sc.et.models.incompatible.obs} and our inferred values are quite reasonable. At $\lesssim 10\,$kpc, the effective diffusivities/streaming speeds at $\sim$\,GeV are in fact quite close to the empirically-favored LISM values (from direct calibration to Voyager, AMS, and other CR experiments with classic models like GALPROP, DRAGON, USINE, etc.) $\kappa_{\rm eff} \sim 1-3\times 10^{29}\,{\rm cm^{2}\,s^{-1}}$ \citep[see e.g.][]{genolini:2019.usine.fits.kappa.vs.cr.diffusion.delta.half.plus.nu,delaTorre:2021.dragon2.methods.new.model.comparison,korsmeier:2021.light.element.requires.halo.but.upper.limit.unconfined,korsmeier:2022.cr.fitting.update.ams02,hopkins:cr.multibin.mw.comparison,dimauro:2023.cr.diff.constraints.updated.galprop.very.similar.our.models.but.lots.of.interp.re.selfconfinement.that.doesnt.mathematically.work}. For the outer halo ($\sim 100\,$kpc), constraints are much more sparse (and there are of course no ``direct'' CR measurements), but Fig.~\ref{fig:kappa} shows that for MW-mass galaxies, the implied $\kappa_{\rm eff}$ agrees  well with the empirical constraints from \citet{butsky:2022.cr.kappa.lower.limits.cgm} around MW-mass galaxies. What is remarkable is that the latter is derived from an entirely independent technique and dataset: using indirect arguments about pressure balance in the CGM combined with UV absorption line spectra from background quasars dominated by cool ($< 10^{5}\,$K) gas (indeed primarily HI absorption, so neutral gas at $\sim 10^{4}\,$K) -- with completely negligibly thermal X-ray emission -- to constrain $\sim$\,GeV proton diffusivities. These are also similar to plausible streaming speeds in theoretical models \citep{wiener:cr.supersonic.streaming.deriv,wiener:2017.cr.streaming.winds,Rusz17,holguin:2019.cr.streaming.turb.damping.cr.galactic.winds,kempski:2021.reconciling.sc.et.models.obs,thomas:2022.self-confinement.non.eqm.dynamics,ponnada:2023.synch.signatures.of.cr.transport.models.fire}.

\subsection{Associated Radio and $\gamma$-ray Halos}
\label{sec:radio.gamma}

If the observed X-ray emission is indeed non-thermal, it is important to calculate the observed emission in other channels. First consider the radio (synchrotron) emission. This will be doubly-suppressed by weak magnetic ($B$) fields in the outer halo and IGM \citep{ponnada:fire.magnetic.fields.vs.obs,ponnada:2023.fire.synchrotron.profiles}, $B\sim 30\,B_{\rm 30}\,{\rm nG}$ \citep{dolag:2011.lower.limit.igm.bfields,vernstrom:2019.40.nanogauss.igm.field.limits.from.rms,temsland:2023.lower.limit.igm.bfields}.\footnote{At $R\sim 100\,$kpc around MW and M31-mass galaxies, cosmological magnetohydrodynamic+galaxy formation simulations predict a range $\langle |{\bf B}|^{2}\rangle^{1/2} \sim (10-100)\,$nG \citep{2015MNRAS.453.3999M,2019MNRAS.484.2620K,ponnada:fire.magnetic.fields.vs.obs,mannings:2023.cgm.bfield.obs.models}, with observational upper limits around such galaxies from Faraday rotation at $R\sim 100\,$kpc of $B < 200-1000\,$nG \citep{prochaska:2019.weak.magnetization.low.Bfield.rm.massive.gal.frb,prochaska:2019.frb.halo.constraints,lan:2020.cgm.b.fields.rm} and at $R \sim $\,Mpc upper limits are $B < (2-20)\,$nG \citep{ravi:2016.igm.b.upper.limits.frb,vernstrom:2019.40.nanogauss.igm.field.limits.from.rms,osullivan:2020.nanoGauss.upper.limits.galactic.B.at.mpc,padmanabhan:2023.igm.bfield.constraints.frbs}.} The bolometric synchrotron luminosity will be suppressed relative to the IC by the ratio $e_{\rm B}/e_{\rm CMB} \sim 7\times 10^{-5}\,B_{30}^{2}\,(1+z)^{-4}$, and the characteristic frequency $\nu_{\rm cr} = (3/2) \gamma^{2}\nu_{\rm G} \sin{\alpha} \sim 0.4\,{\rm MHz}\,B_{30}\,E_{\rm cr,\,GeV}^{2}$. So powering $1.4$\,GHz emission, for example, requires $E_{\rm cr,\,GeV} \sim 60\,B_{30}^{-1/2}$ -- i.e.\ extremely high-energy electrons where the CR number density is strongly suppressed even assuming an LISM spectrum (let alone correcting for IC losses in the halo). {\em Neglecting} those losses or simply parameterizing them with some term $f_{\rm loss}^{\sim 100\,{\rm GeV}} \ll 1$, we obtain a $1.4\,$GHz emissivity $\epsilon_{1.4\,{\rm GHz}} \sim 10^{29.3}\,{\rm erg\,s^{-1}\,kpc^{-3}}\,B_{30}^{9/4}\,(f_{\rm loss}^{\sim 100\,{\rm GeV}}\,e_{\rm cr,\,\ell}/{\rm 0.02\,eV\,cm^{-3}})$, and flux 
$S_{1.4\,{\rm GHz}} \sim 10^{31.5}\,{\rm erg\,s^{-1}\,kpc^{-2}}\,B_{30}^{9/4}\,R_{100}\, f_{\rm loss}^{\sim 100\,{\rm GeV}}\, \times (e_{\rm cr,\,\ell}/{\rm 0.02\,eV\,cm^{-3}})$, 
or integrated luminosity
\begin{align}
\frac{L_{1.4\,{\rm GHz}}}{\rm W\,Hz^{-1}} &\sim 10^{17.8}\,B_{30}^{9/4}\,R_{100}^{3}\,
\left( \frac{e_{\rm cr,\,\ell}\,f_{\rm loss}^{\sim 100\,{\rm GeV}}}{\rm 0.0002\,eV\,cm^{-3}}\right)\,  , \\ 
\nonumber &\sim 10^{18.2}\,f_{\rm loss}^{\sim 100\,{\rm GeV}}\,B_{30}^{9/4}\,L^{\rm tot}_{\rm X,\,keV,\,40}\ , 
\end{align}
about $\sim 3-4$\,dex fainter than even the lowest-luminosity detected synchrotron galaxy disks/bulges \citep{magnelli:2015.fir.radio,delhaize:2017.fir.radio,wang:2019.fir.radio.corr} even if $f_{\rm loss}^{\sim 100\,{\rm GeV}}=1$ (no losses). To be detectable from $R\sim 100\,{\rm kpc}$ to $R\sim R_{\rm vir}$ in terms of either surface brightness or luminosity, even within distances $<10\,$Mpc (let alone the X-ray sample distances which are at cosmological redshifts $z\sim 0.05-0.2$, i.e.\ luminosity distances $\sim 250-1000\,$Mpc), these would require $B \gtrsim {\rm few}\,{\rm \mu G}$, $e_{\rm cr,\,\ell} \gtrsim {\rm 0.02\,eV\,cm^{-3}}$, and $f_{\rm loss}^{\sim 100\,{\rm GeV}}\sim1$ i.e.\ ISM-level magnetic field strengths  and CR energy densities without any IC losses. These are orders-of-magnitude larger than any realistic model would predict at $>100\,$kpc distances from the central galaxies, except in massive clusters where $100\,{\rm kpc} < 0.1\,R_{\rm vir}$ and these numbers are plausible (precisely where radio synchrotron halos are indeed observed; \citealt{gitti:2016.radio.minihalos.coolcore.clusters.candidates.review}). And again, this is an {\em upper limit} neglecting losses: because $1.4$\,GHz is so sensitive to high-energy leptons, the fluxes are reduced (for the same {\em total} leptonic energy $e_{\rm cr,\,\ell}$) by an additional $\sim 5$\,dex ($f_{\rm loss}^{\sim 100\,\rm GeV} \sim 10^{-5}$) for even a modest IC correction (unless one could observe at much lower $\lesssim $\,MHz frequencies, tracing the GeV leptons).

Because of this suppression, and the further, additional suppression factor by the scattering rates above and $n_{\rm photons,synch}/n_{\rm photons,\,CMB}$, it follows that synchrotron self-Compton contributes negligibly in the X-rays. And owing to the low gas densities in the outer halos, the non-thermal CR bremsstrahlung emission is also extremely faint \citep[c.f.\ scalings in][]{blumenthal:1970.cr.loss.processes.leptons.dilute.gases}. 

If we assume an LISM-like proton spectrum and proton-to-electron ratio, the $\sim$\,GeV $\gamma$-ray emissivity (accounting for all the salient hadronic interactions, appropriate branching ratios, and spectrum at $\sim 0.1-10\,$GeV energies) is $\approx 1.8 \times 10^{-28}\,{\rm erg\,s^{-1}\,cm^{-3}}\,(e_{\rm cr,\,tot}/{\rm eV\,cm^{-3}})\,(n_{\rm gas}/{\rm cm^{-3}})$ or $5.3 \times 10^{31}\,{\rm erg\,s^{-1}\,kpc^{-3}}\,(e_{\rm cr,\,tot}/{\rm eV\,cm^{-3}})\,(n_{\rm gas}/10^{-5} {\rm cm^{-3}})$ \citep{guo.oh:cosmic.rays,lacki:2011.cosmic.ray.sub.calorimetric,wiener:cr.supersonic.streaming.deriv,hopkins:cr.transport.constraints.from.galaxies}, so (assuming the typical profiles above) $S_{\gamma} \sim 8\times 10^{33}\,{\rm erg\,s^{-1}\,kpc^{-2}}\,(e_{\rm cr,\,tot}/{\rm eV\,cm^{-3}})\,(n_{\rm gas}/10^{-5}\,{\rm cm^{-3}})\,R_{100}$ and 
\begin{align}
\nonumber \frac{L_{\gamma}}{\rm erg\,s^{-1}} &\sim 2\times10^{36}\, R_{100}^{3}
\left( \frac{e_{\rm cr,\,tot}}{\rm 0.01\,eV\,cm^{-3}}\right) \left( \frac{n_{\rm gas}}{\rm 10^{-5}\,cm^{-3}}\right)\, , \\ 
 & \sim 
6\times10^{36}\,\left(\frac{f_{\rm cr,\,\ell}^{\odot}}{f_{\rm cr,\,\ell}}\right) \left( \frac{n_{\rm gas}}{\rm 10^{-5}\,cm^{-3}}\right) L^{\rm tot}_{\rm X,\,keV,\,40}\, , 
\end{align}
where we have scaled to a typical gas density seen in simulations at virial radii \citep{Buts18,ji:fire.cr.cgm,ji:20.virial.shocks.suppressed.cr.dominated.halos,ramesh:2024.tng.plus.our.subgrid.crs.very.strong.fb.fx}.
For comparison, the sensitivity of {\em Fermi} is such that outside of $<1\,$Mpc, the lowest-luminosity detectable systems (NGC253 and M82, both in the local group) have $L_{\gamma} > 10^{40}\,{\rm erg\,s^{-1}}$ \citep{lacki:2011.cosmic.ray.sub.calorimetric,lopez:2018.smc.below.calorimetric.crs,zhang:2019.new.cosmic.ray.compilation.vs.calorimetry.sub.calor}, and for systems in the same range of distances/redshifts as the ROSAT or eROSITA samples, detection typically requires $L_{\gamma} \gg 10^{41}-10^{42}\,{\rm erg\,s^{-1}}$ \citep{ackermann:2014.cosmic.ray.fermi.gamma.ray.upper.limits.galaxy.clusters.data.not.as.model.dependent}. 
So the extended $\gamma$-ray halo would be completely undetectable unless, again, we assumed an un-realistically large $e_{\rm cr,\,tot}$ and $n_{\rm gas}$ for MW-Andromeda like host halos out to even larger radii. Once again, though, the predictions do become interesting for the central regions of massive clusters, where non-trivial constraints from $\gamma$-rays do exist \citep[e.g.][]{ackermann:2014.cosmic.ray.fermi.gamma.ray.upper.limits.galaxy.clusters.data.not.as.model.dependent}, although these would need to be revisited in an IC+CR-dominated scenario (since the gas densities may be very different from what was assumed). 

\begin{figure}
	\centering\includegraphics[width=0.99\columnwidth]{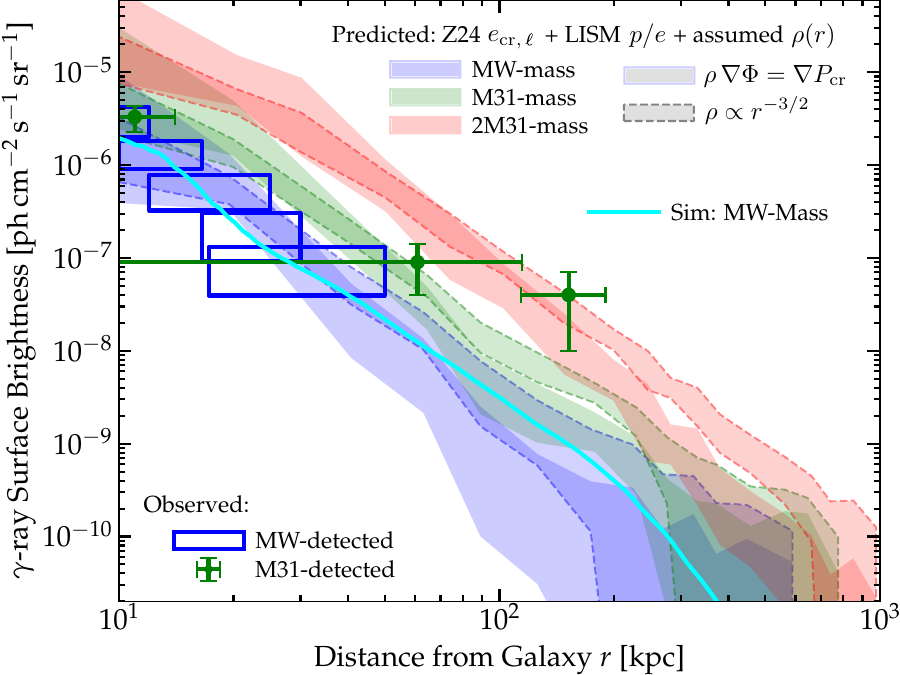} 
	\centering\includegraphics[width=0.99\columnwidth]{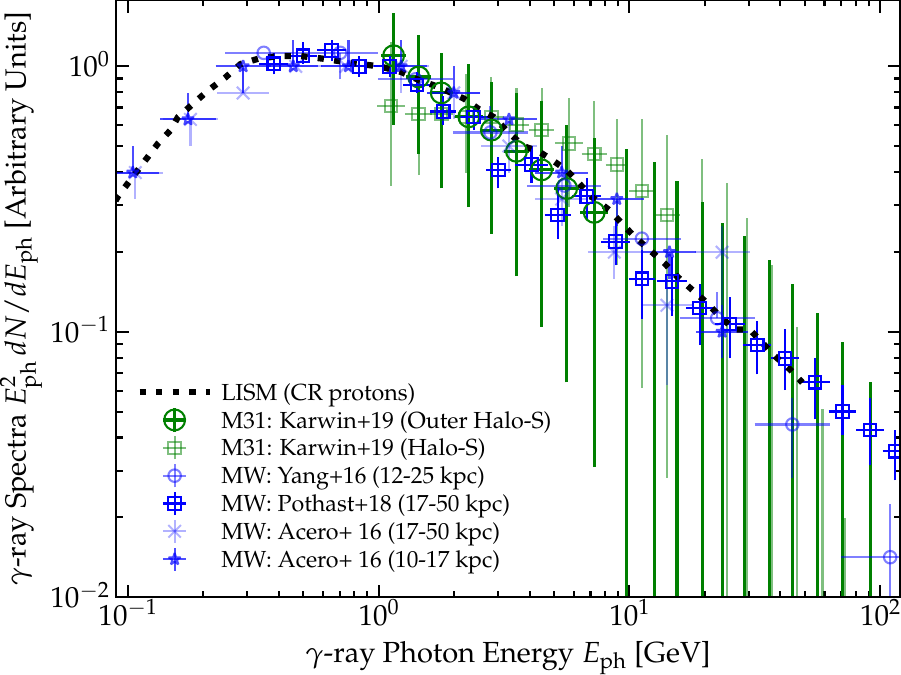} 
	\caption{{\em Top:} Associated $\gamma$-ray emission ({\em shaded}) from the CRs expected to explain the observed soft X-rays (Fig.~\ref{fig:ecr}), assuming (1) an LISM electron-to-proton ratio and proton spectrum \citep{bisschoff:2019.lism.cr.spectra}; and (2) a gas density profile either (a) in hydrostatic equilibrium with CR pressure alone ({\em solid}) $\nabla P_{\rm cr} = \rho\,\nabla \Phi$ (with $\Phi$ given by an NFW profile for the sample median halo mass in Z24), or (b) containing $1/2$ of the universal baryon fraction out to $R_{\rm vir}$ with a power-law $\rho \propto r^{-3/2}$ ({\em dashed}). We compare extended diffuse CGM emission detected at different radii in the MW \citep{acero:2016.gamma.ray.constraints.cr.emissivity,yang.2016:diffuse.gamma.ray.cr.profile.constraints,pothast:2018.radial.mw.gamma.profile.modeling.fermi} and M31 \citep{abdo:2010.outer.mw.gamma.ray.emission,karwin:2019.fermi.m31.outer.halo.detection}. 
	We also compare the prediction from the cosmological MW-mass galaxy simulation in Fig.~\ref{fig:profiles}.
	{\em Bottom:} Observed $\gamma$-ray spectra from {\em Fermi} in the different MW and M31 regions detected above, compared to the LISM $\gamma$-ray spectrum from CRs in diffuse gas (dominated by pion production from LISM CR protons). The observed spectra and profiles are completely consistent with the LISM CR proton spectrum ``leaking'' out of galaxies at the expected rate. Per \S~\ref{sec:radio.gamma}, these would be undetectable in galaxies beyond M31. 
	\label{fig:gamma.profile}}
\end{figure}

There are of course two systems of interest {\em within} $<1\,$Mpc, namely the MW and M31, where some diffuse $\gamma$-ray emission has been detected. In Fig.~\ref{fig:gamma.profile} we show the profiles measured by Fermi from ring analysis and windows constraining the outer Galaxy \citep{abdo:2010.outer.mw.gamma.ray.emission,ackermann.2011:diffuse.gamma.ray.cr.profile.constraints,tibaldo.2014:diffuse.gamma.ray.cr.profile.constraints,tibaldo.2015:diffuse.gamma.ray.cr.profile.constraints,tibaldo.2021:diffuse.gamma.ray.cr.profile.constraints,acero:2016.gamma.ray.constraints.cr.emissivity,yang.2016:diffuse.gamma.ray.cr.profile.constraints,pothast:2018.radial.mw.gamma.profile.modeling.fermi},\footnote{Compiled in \citet{hopkins:cr.multibin.mw.comparison}, using the median $N_{\rm HI}$ in those maps to convert from $\gamma$-ray emissivity to surface brightness where necessary. Note these studies also give consistent scalings for the emissivity to what we adopt (within a range much smaller than uncertainties in Fig.~\ref{fig:gamma.profile}).} and the detections of M31 both on ISM scales and the ``spherical/outer halo'' diffuse emission (\citealt{abdo:2010.outer.mw.gamma.ray.emission,karwin:2019.fermi.m31.outer.halo.detection}; see also \citealt{recchia:2021.gamma.ray.fermi.halos.around.m31.modeling,do:2021.cr.transport.m31.halo.modeling}).\footnote{The outer radii constraints are unaffected by the central source model discussion in \citet{yi:2023.m31.two.point.sources.in.center}. For the spectra in Fig.~\ref{fig:gamma.profile}, we follow \citet{do:2021.cr.transport.m31.halo.modeling} and focus on the M31 ``S'' window which \citet{karwin:2019.fermi.m31.outer.halo.detection} argue is less affected by foreground contamination.} Note the large range of radii in both cases. We compare these to the predicted $\gamma$-ray emission (1) from the same simulation as Fig.~\ref{fig:profiles}, which features a CR-pressure dominated halo; (2) assuming an LISM proton-to-electron ratio (i.e.\ same as used to estimate CR pressure below in Fig.~\ref{fig:Pcr}) and LISM proton spectra and heuristic density profile matching the simulation (assuming the diffuse halo contains $50\%$ of the universal baryon fraction within $R_{\rm vir}$, with a power-law $n_{\rm gas}\propto r^{-3/2}$ profile); or (3) assuming again LISM proton spectra but calculating the gas density $n_{\rm gas}$ which would be in quasistatic equilibrium supported by CR pressure alone.\footnote{Specifically following \citet{butsky:2022.cr.kappa.lower.limits.cgm}, assuming a spherical halo with $\nabla P_{\rm cr} = \rho_{\rm gas}\,\nabla \Phi = -\rho_{\rm gas}\,V_{c}^{2}/r$, where we assume the circular velocity $V_{c}$ is given by a dark matter halo with an NFW profile, concentration $c_{\rm vir}\sim10$, and halo mass from \citet{zhang:2024.hot.cgm.around.lstar.galaxies.xray.surface.brightness.profiles} for each sample, while $P_{\rm cr}$ follows Fig.~\ref{fig:Pcr}.} We stress these are just assumptions -- the X-rays do not directly constrain these ratios. Nonetheless, it is striking that the predicted $\gamma$-ray profiles assuming LISM-like CR $p/e$ and spectra and a CR-pressure dominated halo appear to be completely consistent with the claimed Fermi detections of diffuse emission from $\sim 10-150\,$kpc in the MW and M31.

$\gamma$-rays can also be produced from IC scattering of higher-energy cosmic IR, optical, UV, and X-ray background photons, but these should be a small correction in the CGM (unlike e.g.\ very near-source and strongly illuminated Galactic center regions or supernova remnants; \citealt{porter:2006.ic.galactic.snrs.isrf.effects,porter:2008.galactic.ridge.cr.ic.xray.gamma.ray.emission.multiprocess}). Integrating over the observed cosmic background spectrum at $z\approx 0$ from $\sim 100\,{\rm \mu m}$ to $\sim 100\,$keV \citep{younger:ir.background,ueda:2014.xray.luminosity.functions.update,cafg:2020.uv.background}, even if we ignore losses (i.e.\ adopt a strictly Solar-neighborhood CR lepton spectrum) we would obtain a $\gamma$-ray emissivity in a broad band around $\sim 0.3-3\,$GeV of $\sim 1.5 \times10^{32} {\rm erg\,s^{-1}\,kpc^{-3}}\,(e_{\rm cr,\,\ell}/{\rm eV\,cm^{-3}})\,f_{\rm loss}^{\sim 30\,{\rm GeV}}$, so this could at most boost the $\gamma$-ray luminosities by a factor of a couple (still leaving the halos undetectable), primarily from IC of the cosmic optical background via $> 30\,$GeV electrons (see e.g.\ \citealt{ackermann:2012.fermi.gamma.rays.ism.cr.emission.modeling} for more detailed calculations of the full $\gamma$-ray spectrum, which give the same conclusion for the MW points in Fig.~\ref{fig:gamma.profile}). But, like with synchrotron, since this is sensitive to much higher energies, we cannot ignore losses: even the conservative loss corrections lower the emissivity by a factor of $\sim 200$ (with the $\gamma$-rays primarily coming from IC of the UVB off few-GeV electrons), leaving this a small correction to the hadronic $\gamma$-rays.

Note that, while we expect the radio and $\gamma$ ray halos to be undetected outside the MW and M31, non-detections are still interesting, as they directly place non-trivial upper limits on magnetic field strengths $B$ and gas densities $n_{\rm gas}$, respectively.

\section{Comparison to Hot Gas Emission}
\label{sec:thermal}

We now compare to what we would obtain if we made the common assumption that X-ray halos arise instead from hot gas emission. We stress that essentially every paper explicitly modeling the eROSITA signal from simulations (references in \S~\ref{sec:intro} and below), as well as the observational papers themselves (both eROSITA, \citealt{zhang:2024.erosita.hot.cgm.around.lstar.galaxies.detected.and.scaling.relations,zhang:2024.hot.cgm.around.lstar.galaxies.xray.surface.brightness.profiles}, and earlier ROSAT, \citealt{anderson:2013.rosat.extended.cgm.xray.halos}), has assumed that the X-ray halos arise from hot gas emission (indeed, titling the papers ``The hot CGM...'' and ``Extended hot halos...'', without mentioning CR-IC). Not only does this overlook potentially interesting processes, but these studies have generally found it leads to significant tensions, which we review here as they are important to interpretation and breaking degeneracies.

\subsection{Spectrum}
\label{sec:thermal.spectrum}

Fig.~\ref{fig:spectrum} compares the predicted IC spectra to pure thermal emission computed with APEC \citep{smith:2001.apec.methods}, for the upper limits to the metallicities at $\sim R_{\rm vir}$ discussed below, including free-free ($d \epsilon_{\rm ff} / d\ln{E_{\rm ph}} \sim 4 \times 10^{30} {\rm erg\,s^{-1}\,kpc^{-3}}\,(n_{\rm gas}/10^{-5}\,{\rm cm^{-3}})^{2} (T/10^{6}\,K)^{1/2} x_{\rm ff} \exp{[-x_{\rm ff}]}$ with $x_{\rm ff} \equiv E_{\rm ph}/k_{B} T$; \citealt{rybicki.lightman:1979.book}) and line emission, similar broadly in continuum shape to the observed spectra of cluster cores. Roughly speaking, the predicted IC spectrum looks similar to a $\sim 3\times10^{7}\,$K free-free-dominated spectrum or a line-dominated $\sim 3\times 10^{6-7}$\,K spectrum, peaking at $\sim$\,keV. 

Even ignoring losses and assuming the spectrum approaches the LISM CR spectrum in the central $\sim 10\,$kpc around the galaxy, the predicted CR-IC spectra are only slightly harder at $\sim 8-10$\,keV, making them effectively indistinguishable from thermal emission in X-ray hardness ratios even if these existed.\footnote{eROSITA $0.2-2$\,keV hardness ratios are published for bright individually-detected sources \citep{nandra:2025.efeds.data}, but the extended CGM-stacked signal is too faint to divide into sub-bands \citep{zhang:2024.erosita.hot.cgm.around.lstar.galaxies.detected.and.scaling.relations,zhang:2024.hot.cgm.around.lstar.galaxies.xray.surface.brightness.profiles}. Likewise hardness ratios from ROSAT are only available stacking together the entire (un-resolved) CGM and central sources, where the latter dominate $L_{X}$  \citep{anderson:2013.rosat.extended.cgm.xray.halos,anderson:2015.rosat.xray.halo.stacking.scaling.relations}.} More common searches for IC emission in bright systems focus on detection of power-law continua at much harder energies, $\sim 20-80\,$keV \citep{fusco:2000.cluster.hardx.ic.detection.weak.B,gruber:2002.rxte.cluster.ic.B.weak,rephaeli:2003.cluster.ic.detection.B.limits.low,bonamente:2007.abell.3112.clear.xray.inverse.compton.required.luminosity.fits.models.gamma.rays.too,chen:2008.ic.searches.hardxray.B.field.lower.limits.clusters,wik:2011.swift.ic.hardx.upper.limits.clusters.b.lower.point1microg,cova:2019.cluster.ic.upper.limits.B.lower.limits,mirakhor:2022.ic.cluster.detection.B.0pt1microG}. But there is no predicted detection: even ignoring losses, while the luminosity in this band would only be a factor of a few smaller than at $\sim 1\,$keV, the predicted fluxes and surface brightnesses are far lower than the sensitivity of instruments at present. For example, given a typical $L_{X,\,40}$, this predicts a $20-80$\,keV integrated flux for a source at the median redshift of $\sim 0.1$ of $\sim 10^{-16}\,{\rm erg\,s^{-1}\,cm^{-2}}$, about $\sim 5$ orders of magnitude lower than typical upper limits in undetected sources  \citep{bartels:2015.radio.inverse.compton.cluster.minihalo.prospects}. Including realistic losses suppresses this exponentially (i.e.\ makes the spectra effectively thermal, as we showed). And because of the scaling of loss rates/lifetimes from IC and synchrotron, almost all of the IC from harder wavelengths is predicted to arise from the central $\lesssim 10\,$kpc within/around the galaxies, so any hard X-ray IC which might be detected would appear as exclusively ISM emission, not CGM emission associated with the halo.

More noticeable differences in line emission appear at the largest radii, when $\tau_{\rm loss}^{0} \gg 1$ (CR losses are much larger) and the CR-IC luminosity is dropping rapidly, because the softer CR-IC spectra are similar to lower-temperature thermal emission (where line emission becomes more important). But (1) this is by definition only occurring at radii where the predicted CR-IC soft X-ray emission is exponentially falling (so even more extremely low-surface-brightness), (2) this is still degenerate, to some extent, with a mix of temperatures and metallicities and densities, and (3) this requires extremely sensitive, high-resolution X-ray spectroscopy to disentangle. Still, that may be possible with future microcalorimeter missions sensitive to very low surface-brightness emission \citep[e.g.][]{kraft2023lineemissionmapperlem}.

\begin{figure}
	\centering\includegraphics[width=0.99\columnwidth]{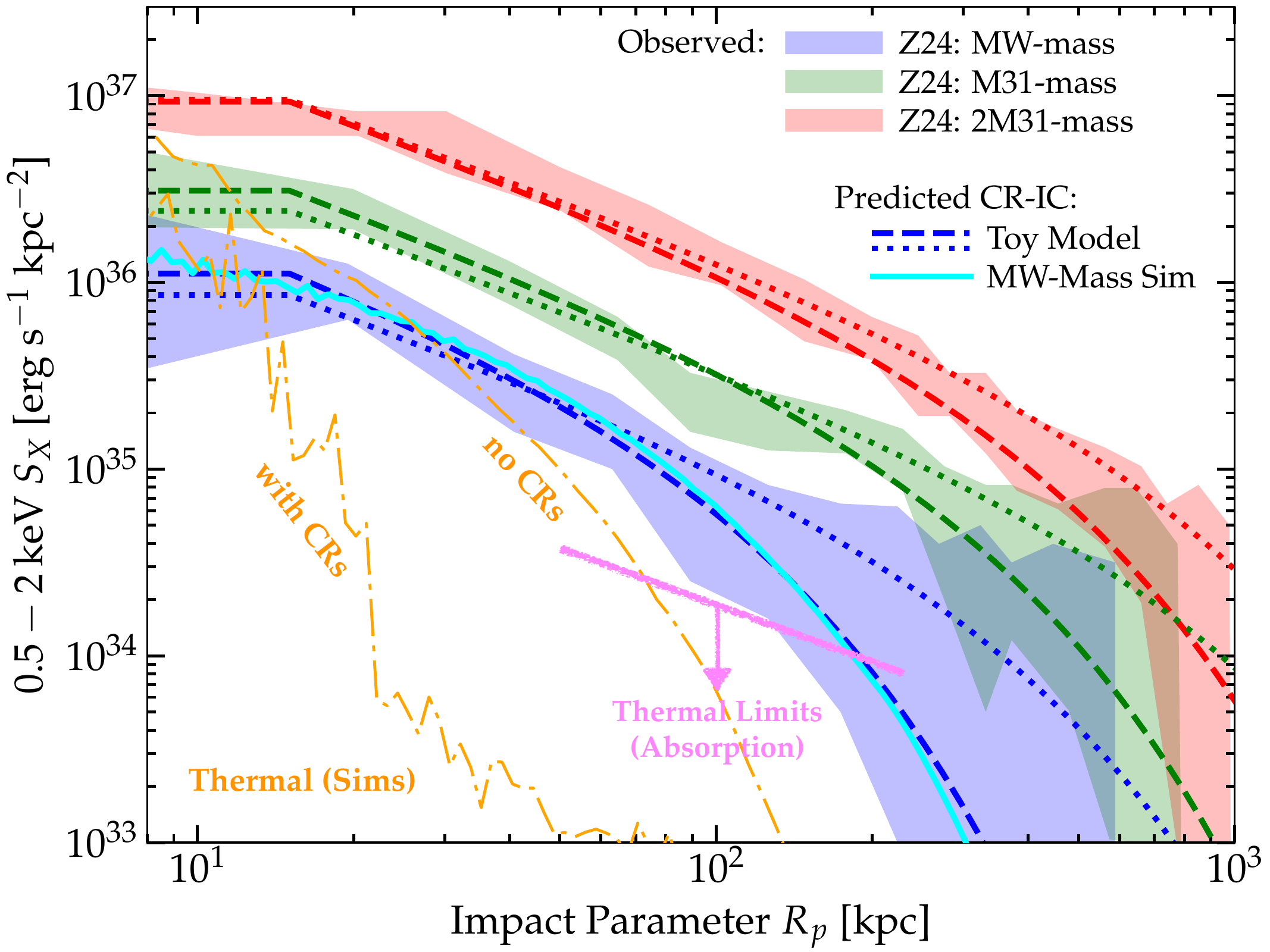} 
	\caption{Soft X-ray emission as Fig.~\ref{fig:profiles}, but comparing the predicted thermal (free-free+line) emission (\S~\ref{sec:thermal.profiles}). 
	We show upper limits to thermal emission implied by X-ray absorption studies with sightlines at impact parameters from $50$ to $\sim250\,$kpc around MW+M31-mass systems.
	We also show the predicted thermal emission from the same numerical CR-MHD cosmological simulation of a MW-mass galaxy, as well as the same galaxy re-simulated with identical physics except no CRs (which produces a significant higher SFR and more massive, M31-mass, galaxy, with more associated thermal emission; \citealt{hopkins:cr.mhd.fire2,ji:fire.cr.cgm}). The latter is similar to predicted thermal emission from other simulations \citep[e.g.\ TNG, EAGLE][]{chadayammuri:2022.compare.sims.xray.cgm.profiles.vs.mass} without CRs.
	\label{fig:profiles.thermal}}
\end{figure}

\subsection{Surface Brightness Profiles}
\label{sec:thermal.profiles}

\subsubsection{Basic Scalings \&\ Simulation Predictions}
\label{sec:thermal:pred}

With that in mind, consider the radial profile of broadband-integrated emission as observed. 
While models we discuss use more detailed calculations \citep[e.g.][]{smith:2001.apec.methods}, for deriving simple scalings and intuition we can roughly approximate the thermal-only $0.5-2$\,keV emissivity of hot gas as the sum of the band-integrated free-free 
$\epsilon_{X,\,{\rm ff}}$ 
(approximated by 
$\epsilon_{X,\,{\rm ff}}/({\rm erg\,s^{-1}\,kpc^{-3}}) \sim 4 \times 10^{30}\,(n_{\rm gas}/10^{-5}\,{\rm cm^{-3}})^{2} (T/10^{6}\,K)^{1/2} x_{\rm ff}^{\rm keV} \exp{[-x_{\rm ff}^{\rm keV}]}$, with $x_{\rm ff}^{\rm keV} \approx {\rm keV}/k_{B}T$; \citealt{rybicki.lightman:1979.book}), 
and lines $\epsilon_{X,\,{\rm line}}$ 
(approximated by 
$\epsilon_{X,\,{\rm line}}/({\rm erg\,s^{-1}\,kpc^{-3}}) \sim (Z/Z_{\odot})\,10^{32}\,(n_{\rm gas}/10^{-5}\,{\rm cm^{-3}})^{2} q^{2}\,\exp{(-q)}/(q+1)$ with $q\equiv 0.3\,{\rm keV}/k_{B}T$; \citealt{mewe:1985.xray.thermal.collisional.line.emissivities}).\footnote{Equivalent to a line-cooling-function $\Lambda_{X}^{0.5-2} \sim 4\,(Z/Z_{\odot}) \times 10^{-23}\,q^{2}/(q+1)\,\exp{(-q)}$.}
As is well-known, these give line emission (for $Z \gtrsim 0.1\,Z_{\odot}$) dominating over free-free at $T \lesssim 5 \times 10^{6}\,$K. 
Given that the virial temperature of halos (at $z \sim 0$, where observed) scales as 
$T_{\rm vir} \approx 5 \times 10^{5}\,{\rm K}\,(M_{\rm vir}/10^{12}\,M_{\odot})^{2/3}$ \citep{bryan.norman:1998.mvir.definition}, 
lines should dominate the emissivity at $M_{\rm vir} \lesssim 10^{13.5}\,M_{\odot}$, 
with negligible contribution (given the scalings above) from gas with $T\lesssim 10^{6}\,$K. 
So for the halos of interest we expect 
$\epsilon^{\rm therm}_{X} \sim 0.8 \times 10^{30}\,{\rm erg\,s^{-1}\,kpc^{-3}}\,(Z/0.1\,Z_{\odot})\,(n_{\rm hot\,gas}/10^{-5}\,{\rm cm^{-3}})^{2}\,f_{\rm hot}$, 
where $f_{\rm hot}$ is the fraction of hot gas with $T > 10^{6}\,$K (which can be $\ll 1$).\footnote{Note that $f_{\rm hot}$ defined this way means that there should be a very rapid falloff in brightness below $M_{\rm halo} \lesssim 10^{12.5}\,M_{\odot}$, as well.} Integrating along a line-of-sight at impact parameter $R$, this gives $S^{\rm therm}_{X} \sim 10^{32}\,(Z/0.1\,Z_{\odot})\,(n_{\rm hot}/10^{-5}\,{\rm cm^{-3}})^{2}\,f_{\rm hot}\,(R/100\,{\rm kpc})$ (in ${\rm erg\,s^{-1}\,kpc^{-2}}$).

Both simulations of MW-M31 mass galaxies (\citealt{guedes:2011.cosmo.disk.sim.merger.survival,keres:2011.arepo.gadget.disk.angmom,liang:2016.cgm.nh.profiles,Buts18,ji:fire.cr.cgm,kim:2022.hot.gas.cgm,truong:2023.cosmo.sims.sb.predictions.verylow.larger.vs.erosita.data,ramesh:2023.cgm.tng50.scalings,medlock:2025.warm.igm.camels}; and see also \citealt{braspenning:2024.flamingo.simple.xray.modeling.sims.clusters.dont.reproduce.zdrops.other.cc.features,sultan:2024.cooling.flows.model.fire.hot.cgm.lowermass.galaxies,lehle:2024.simulation.cluster.profiles} for more massive halos) and observations from UV absorption lines in similar mass halos \citep{machado:2018.cgm.sims.vs.obs,burchett:2019.neviii.halo.gas,faerman:2022.cgm.props.needed.obs.vs.sams}, as well as resolved and individually-detected X-ray profiles in more massive groups/clusters \citep{cavagnolo:cluster.entropy.profiles,ettori:2019.hydrostatic.cluster.mass.profiles,ghirardini:2019.cluster.profiles.compilation.universal.fits,luskova:2023.erosita.xray.cluster.surface.brightness.profiles} predict at these radii $n_{\rm hot\,gas} \sim 4\times 10^{-5}\,(r/100\,{\rm kpc})^{-\alpha_{n}}$ with $\alpha_{n}$ between $2$ and $3$, $T \propto r^{-\alpha_{T}}$ with $\alpha_{T} \sim 0.5$, and $Z \sim 0.1\,(r/100\,{\rm kpc})^{-\alpha_{Z}}$ with $\alpha_{Z}$ between $0.5-1$. Inserting these into the scalings above, the predicted thermal-only surface brightness scales as $S_{X}^{\rm therm} \sim 10^{33}\,{\rm erg\,s^{-1}\,kpc^{-2}}\,f_{\rm hot}\,(R/100\,{\rm kpc})^{-\alpha_{R}}$ with $\alpha_{R}$ between $3.5$ and $6$ (and indeed, in said clusters, $\alpha_{R}$ is typically {\em observed} between $4$ and $5$, as expected). 

Two problems are evident (see Fig.~\ref{fig:profiles.thermal}): (1) the normalization of $S_{X}$ at $\gtrsim 100$\,kpc is predicted to be $\sim 100 \times$ lower than observed in the low-mass halos with virial masses $10^{12.5-13}\,M_{\odot}$ (the discrepancy increasing by another order-of-magnitude for halos with masses $\sim 10^{12}\,M_{\odot}$, owing to the temperature dependence above), and (2) the {\em slope} of $S_{X}$, namely $\alpha_{R}$, is predicted to be qualitatively different, with a very steep $\propto R^{-4.5}$ approaching $R_{\rm vir}$ as opposed to the observed very shallow $\propto R^{-1}$. Note that if we assume thermal free-free dominates instead (i.e.\ $Z$ very small), both problems become more severe.

These simple scalings help explain why almost every cosmological simulation study which has explicitly forward-modeled the thermal soft X-ray emission (with detailed APEC/XSPEC-type models) from the CGM around MW-mass galaxies features both of the problems above. This includes (but is not limited to) EAGLE, Illustris, FLAMINGO, TNG, SIMBA, Astrid, Magneticum, and FIRE simulations \citep[see e.g.][]{vandevoort:sz.fx.hot.halos.fire,chadayammuri:2022.compare.sims.xray.cgm.profiles.vs.mass,wijers:2022.xray.line.emission.cgm.modeling.eagle,truong:2023.cosmo.sims.sb.predictions.verylow.larger.vs.erosita.data,lau:2024.erosita.profiles.require.cosmological.constraints.violations,zuhone2024propertieslineofsightvelocityfield,vladutescu:2025.magneticum.erosita.profiles.xrb.contributions}, spanning wildly different numerical methods and physical treatments of ``feedback'' from stars and black holes. But notably, none of these papers considers (nor even mentions) IC as an emission mechanism.

\subsubsection{Intrinsic Observational Tensions: UV \&\ X-Ray Absorption}
\label{sec:thermal:obs}

Explaining the observed $S_{X}$ via hot gas emission is not just problematic in terms of comparisons to simulations or more massive observed systems. 
Given the scalings in \S~\ref{sec:thermal:pred}, if we wish to fit the observed $S_{X}$ at large radii with hot gas emission, we require $n_{\rm hot} \sim 10^{-4}\,(M_{\rm vir}/10^{12.5}\,M_{\odot})^{1/2}\,{\rm cm^{-3}}\,(100\,{\rm kpc}/r)\,(Z_{\odot}/Z)^{1/2}\,f_{\rm hot}^{-1/2}$. Integrating this out to $R_{\rm vir}$, this would predict a baryonic mass larger than the Universal baryon fraction times the virial mass, unless $f_{\rm hot}$ and $Z$ are nearly constant with $f_{\rm hot} \sim 1$ and $Z \sim Z_{\odot}$ (as shown also in \citealt{truong:2023.cosmo.sims.sb.predictions.verylow.larger.vs.erosita.data,zhang:2024.hot.cgm.around.lstar.galaxies.xray.surface.brightness.profiles}\footnote{Note that \citet{zhang:2024.hot.cgm.around.lstar.galaxies.xray.surface.brightness.profiles} use an incorrect expression for the virial temperature of their low-mass halos which is a factor of $\sim 5$ higher than the standard expressions and those given and tested in adiabatic simulations in \citet{bryan.norman:1998.mvir.definition} which they used to define $M_{\rm vir}$ and $R_{\rm vir}$. They also do not include the full baryonic mass of the inner galaxy in their available budget. Correcting for this, the  baryonic masses they require to explain the observed $S_{X}$ via thermal emission go up by a factor of $\sim 4-5$, making the tension with cosmological constraints much more severe.}). 

But that, in turn, leads to other contradictions. 
First, this requires the implausible assumption for $M_{\rm vir} \lesssim 10^{13}\,M_{\odot}$ that almost all the gas in the halo is unbound ($T \gg T_{\rm vir}$, $|\nabla P_{\rm gas} | \gg |\rho\,G\,M_{\rm halo}/R^{2}|$), despite no gas being lost (having nearly the entire Universal baryon fraction).
Second, the total metal mass of the halo would be an order of magnitude larger than the maximum possible assuming every SNe had uniformly mixed its metals throughout the halo out to $R_{\rm vir}$, without losses (this predicts a maximum metallicity\footnote{For example, at MW mass, assuming the entire Fe yield of every core-collapse and Ia SN (since Fe is what matters for the calculation here) integrated over the history of the galaxy \citep{maoz:Ia.rate} was retained and uniformly mixed into the halo (with a universal baryon fraction), would predict [Fe/H]$\,\approx-1$ using the most optimistic yields from \citet{leung.nomoto:2018.Ia.yield.model.update}.} of $\sim 0.1\,Z_{\odot}$ at $R_{\rm vir}$, at these halo masses; see \citealt{ma:2015.fire.mass.metallicity,muratov:2016.fire.metal.outflow.loading}). 

This intrinsic challenge is highlighted recently in \citet{lau:2024.erosita.profiles.require.cosmological.constraints.violations}, who use CAMELS to survey a large range of simulation models (TNG, SIMBA, Astrid) and feedback parameters for each model, and show that the only models which can reproduce the eROSITA $S_{X}$ profiles not only require order-of-magnitude larger feedback energetics, but produce a stellar-mass halo mass ($M_{\ast}-M_{\rm halo}$) relation and (equivalently) stellar mass function vastly different from observed (discrepant at $\sim 20-50\sigma$ significance). Interestingly, the way these models ``fit'' the eROSITA signal is not simply by heating gas more effectively with stronger feedback (which is limited by the fact that if gas it heated too much, it simply ``blows out'' of the halo), but shifting $M_{\ast}-M_{\rm halo}$ such that MW-stellar mass galaxies are placed in much more massive halos ($\gtrsim 10^{13.5}\,M_{\odot}$), where the the total gas mass and virial temperature are order-of-magnitude larger.

Regardless of these physical limits (also illustrated in Fig.~\ref{fig:profiles.thermal}), assuming $f_{\rm hot} \sim 1$ and $Z \sim Z_{\odot}$ also contradict what is known from UV and X-ray absorption studies of these halo masses. UV absorption-line observations imply $Z \lesssim 0.1\,Z_{\odot}$, as expected, and $f_{\rm hot} \ll 1$ for $\sim 10^{12}\,M_{\odot}$ halos, based on the ratios of HI and cooler-to-hotter lines \citep{werk:2014.cos.halos.cgm,faerman:2022.cgm.props.needed.obs.vs.sams,wijers:2024.neviii.failure.of.simulations}. 
Very similar constraints -- $Z\sim 0.05\,Z_{\odot}$, $f_{\rm hot} \lesssim 0.1$ -- are also obtained from eROSITA X-ray absorption studies of the MW halo \citep{ponti:2023.erosita.supervirial.gas.close.to.galaxy.cgm.low.metal.and.low.density}, and from stacked Chandra absorption-line studies of galaxies along quasar sightlines \citep{yao:2010.chandra.upper.limits.warm.hot.cgm.gas.in.mw.mass.halos}.

Some super-virial gas is indeed detected in Galactic X-ray absorption studies \citep[e.g.][]{das:supervirial.gas.external.galaxy.absorption,ponti:2023.erosita.supervirial.gas.close.to.galaxy.cgm.low.metal.and.low.density,bhattacharyya:2023.hot.cgm.mw.gas.shocked.close.to.disk,roy:2024.super.virial.gas.obs.above.plane} and much more massive galaxy groups \citep{lovisari:2021.galaxy.groups.review}, and predicted in simulations \citep{chan:2021.cosmic.ray.vertical.balance,roy:2024.super.virial.gas.near.shocks.in.inner.cgm}. But importantly these observations imply that the Galactic super-virial gas is a  small fraction of the Universal baryon fraction ($\sim 10^{8}\,M_{\odot}$, as opposed to $\sim {\rm few} \times 10^{11}\,M_{\odot}$) and that it is  concentrated near the halo center (at $R \lesssim 20\,$kpc). A volume-filling component with $f_{\rm hot} \sim 1$ and $Z\gtrsim 0.3\,Z_{\odot}$ at $\sim R_{\rm vir}$ would be easily detected by eROSITA and Chandra in absorption, and is strongly ruled-out.

\subsubsection{Caveats and Total vs.\ CGM Luminosities}
\label{sec:caveats}

Again we stress the tensions in \S~\ref{sec:thermal:pred}-\ref{sec:thermal:obs} apply to low-mass (MW, M31, and smaller) systems ($M_{\ast} \ll 2\times10^{11}\,M_{\odot}$). At the massive group scale and above, the scalings above predict more thermal emission and there is no problem reproducing the observed profiles at $\sim R_{\rm vir}$ (as is well-known, see references in e.g.\  \citealt{braspenning:2024.flamingo.simple.xray.modeling.sims.clusters.dont.reproduce.zdrops.other.cc.features,lehle:2024.simulation.cluster.profiles}). 

We also caution that the majority of historical comparisons to simulations at the lower mass end focus on just the $L_{X}-M_{\ast}$ relation \citep[][and references therein]{popesso:2024.erosita.stacking.lx.mhalo.by.halo.mass}. This is in principle much easier to reproduce, as one could conceivably generate the observed total $L_{X}$ with a relatively small ($\sim 10^{8}\,M_{\odot}$) mass of Solar-metallicity gas concentrated near the halo center ($R\sim 10\,$kpc) at much higher densities $n_{\rm gas} \sim 10^{-2}\,{\rm cm^{-3}}$, shock-heated to super-virial temperatures $\sim 5 \times 10^{6}\,$K near the peak efficiency of keV emission (which would give $L_{X,\,40}^{Z=Z_{\odot},\,T=T_{\rm peak}} \sim (n_{\rm gas}/10^{-2}\,{\rm cm^{-3}})^{2}\,(R/{\rm 5\,kpc})^{3}$). And some individual (often starburst) galaxies do have X-ray detections of hot outflows at these near-ISM radii (famously M82, but see also \citealt{li.wang:2013.chandra.survey.inclined.disks.xray.near.disk.outflows,sacchi:2025.xrays.near.ism.nearby.galaxies,he:2025.near.disk.hot.halo.xray.obs.outflows}), and in the MW much lower-surface-brightness near-disk emission \citep[e.g.\ the Fermi/eROSITA bubbles][]{predehl:2020.xray.bubbles.mw} and absorption (\S~\ref{sec:thermal:obs}) are seen. And in many low-mass galaxies in observations \citep{anderson:2013.rosat.extended.cgm.xray.halos,zhang:2024.erosita.hot.cgm.around.lstar.galaxies.detected.and.scaling.relations} and simulations \citep{vandevoort:sz.fx.hot.halos.fire,chan:2021.cosmic.ray.vertical.balance,vladutescu:2025.magneticum.erosita.profiles.xrb.contributions} the total X-ray emission {\em including the galaxy} and CGM together are dominated by XRBs (HMXBs) and/or AGN within the central galaxy ($\lesssim 5\,$kpc). 
More detailed modeling of these regions, involving a mix of radiation sources, can be found in e.g.\ \citet{strong:2010.galprop.multiwavelength.cr.emission.across.wide.range,yang:2022.mw.bubbles.modeling.agn.activity,porter:2024.fire.superbubble.properties.distributions,sarkar:2024.fermi.erosita.bubbles.review}, and connections to the much more extended CGM emission will are discussed in \citet{lu:2025.cr.transport.models.vs.uv.xray.obs.w.cric} and Sands et al., in prep.

The challenge, if we ignore CR-IC emission, is how to reproduce the shallow profiles of $S_{X}$ and emission at $\gtrsim 100\,$kpc in MW-mass and smaller halos without violating other observational and/or cosmological constraints. 
There may still be ways to do so: for example, feedback and yield models could be fundamentally incorrect. Or thermal emission could be boosted by very dense clumps with small volume-filling factor (giving a clumping factor $\langle n^{2} \rangle/\langle n\rangle^{2} \gtrsim 100$) for reasons not captured in the simulations reviewed above. The challenge here would be that the emitting CGM gas would have to be dense and compact, but also have $T \gtrsim T_{\rm vir}$, so would be extremely over-pressurized, and can also not be {\em so} compact that it becomes detectable as point-source-like. Of course, the observations could be incorrect, owing to e.g.\ incorrect point-source subtraction or PSF modeling (though studies like \citealt{chadayammuri:2022.compare.sims.xray.cgm.profiles.vs.mass,vladutescu:2025.magneticum.erosita.profiles.xrb.contributions,shreeram:2025.want.to.fit.erosita.w.illustris.have.to.change.halo.masses.and.renormalize.satellites.and.assume.central.psf.not.subtracted.and.psf.wrong.and.2x.count.satellite.lum.w.central.and.assume.all.at.third.solar.metal} argue this could only be important at $R \lesssim 10-40\,$kpc), or various misclassification or stacking effects (though the effects seen in \citealt{shreeram:2024.obs.projection.effects.misclassification.could.bias.soft.xray.cgm.profiles,popesso:2024.stacking.on.halo.mass.extended.xray.contamination.big.issue} would be insufficient to change our conclusions), or incorrect background subtraction. 
However none of the above scenarios changes our most important point, that IC should not be completely neglected (as it usually has been) as a potential emission mechanism in the CGM of $L^{\ast}$ galaxies.

\section{Implications for CR Pressure In the CGM}
\label{sec:pressure}

\begin{figure}
	\centering\includegraphics[width=0.99\columnwidth]{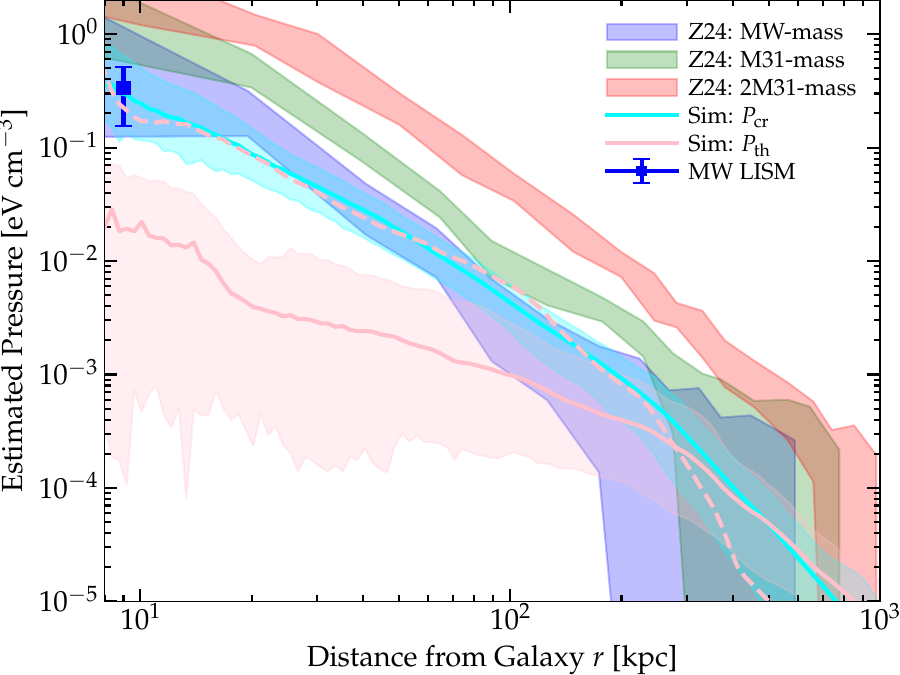} 
	\caption{CR pressure ($P_{\rm cr} \approx e_{\rm cr}/3$) profiles implied by the observed soft X-ray emission from Z24, if we assume an LISM-like electron-to-proton ratio (labeled), motivated by the assumption that GeV electrons and protons both escape the LISM. We compare the directly-measured MW LISM CR pressure ({\em point}). We also compare CR ({\em cyan}) and thermal ({\em pink}) pressure in the same cosmological simulation of a MW-mass galaxy from Fig.~\ref{fig:profiles}, as well as a simulation of the same galaxy with no CRs ({\em pink dashed}) where the pressure is primarily thermal.
	The observed X-ray halos appear to imply that CR pressure plays an important role in the CGM.
	\label{fig:Pcr}}
\end{figure}

Given the inferred $e_{\rm cr,\,\ell}(R)$ from the observations in Fig.~\ref{fig:ecr}, it is natural to ask what this implies for the role of CR pressure ($P_{\rm cr} \sim (1/3)\,e_{\rm cr,\, tot}$, assuming ultra-relativistic particles whose scattering mean-free-path is smaller than the virial radius) in the CGM of MW and M31-mass galaxies. This depends on $e_{\rm cr,\,tot}$, not $e_{\rm cr,\,\ell}$, so is not directly constrained by the X-ray data (except to set a lower limit since $e_{\rm cr,\,tot} \ge e_{\rm cr,\,\ell}$). But since the GeV electrons (which dominate the X-ray signal) in the halo are consistent with what we expect to escape the ISM, where $e_{\rm cr,\,tot}$ is dominated by GeV protons, it is plausible to assume a Solar-neighborhood-like $f_{\rm cr,\,\ell} \equiv e_{\rm cr,\,\ell}/e_{\rm cr,\,tot} \sim f_{\rm cr,\,\ell}^{\odot} \sim 0.01-0.02$.  This assumption is also supported by the fact that CR electrons and protons of the same rigidity (and thus the same energy for $\gtrsim $ GeV)  have the same basic transport physics (diffusion coefficient and/or streaming speed). Since proton losses are negligible, this also supports $f_{\rm cr,\,\ell} \equiv e_{\rm cr,\,\ell}/e_{\rm cr,\,tot} \sim f_{\rm cr,\,\ell}^{\odot} \sim 0.01-0.02$, at least until IC electron losses are important.  The resulting estimated $P_{\rm cr}(R)$ is shown in Fig.~\ref{fig:Pcr}. 

We compare this with the predicted thermal $P_{\rm th} \equiv n k_{B} T$ or CR pressure profiles in cosmological MHD+CR simulations of MW-mass halos from \citet{hopkins:cr.mhd.fire2} with the Feedback In Realistic Environments (FIRE; \citealt{hopkins:2013.fire,hopkins:fire2.methods,hopkins:fire3.methods}) model for galaxy formation and stellar feedback. Specifically, we compare (a) simulations without CRs, where the pressure is almost entirely thermal (there is some small magnetic+turbulent contribution at large radii, but these are not dominant; \citealt{hopkins:cr.mhd.fire2,ji:fire.cr.cgm}), or (b) simulations with CRs using a simple streaming+diffusion model, which form a CR pressure-dominated halo. We see that the observationally-inferred total CR pressure in MW-mass halos agrees well with the total  pressure in either simulation: this necessarily comes from thermal pressure in the ``no CRs'' simulation (a), or from CRs in the CR-dominated simulation (b). This, and the agreement with the MW and M31 $\gamma$-ray data assuming a CR-pressure dominated equilibrium halo in Fig.~\ref{fig:gamma.profile}, strongly suggests that CRs are indeed an important component of the CGM pressure in MW-mass galaxies.\footnote{Note that $P_{\rm cr} \gtrsim P_{\rm therm}$ does not imply CR heating is important relative to thermal cooling. Indeed CR-MHD simulations generally find CR pressure-dominated CGM halos are cooler than halos with $P_{\rm cr} \ll P_{\rm therm}$ \citep{Sale16,Chen16,chan:2018.cosmicray.fire.gammaray,ji:fire.cr.cgm,Buts18,butsky:2020.cr.fx.thermal.instab.cgm,ruszkowski.pfrommer:cr.review.broad.cr.physics,peschken:2022.crs.naab.sims.outflows.similar.to.fire,weber:2025.cr.thermal.instab.cgm.fx.dept.transport.like.butsky.study,farcy:2025.cr.feedback.eor.galaxies.crs.increased.with.sne.decreased.lower.escape.fraction.but.similar.stellar.masses}.}

\section{Conclusions}
\label{sec:conclusions}

Stacked observations have argued that there is detected soft X-ray ($\sim$\,keV) diffuse emission with very shallow profiles $S_{X} \propto R^{-1}$ out to the virial radius around MW and M31-mass galaxies ($M_{\ast} \lesssim 10^{11}\,M_{\odot}$). Attempting to interpret this emission as thermal free-free or metal-line cooling from in-situ gas is challenging: the profiles are far more shallow than simulations or individually-measured thermal emission in more massive halos, factors $\gtrsim 100$ more luminous than generally predicted at $\gtrsim 100\,$kpc, and attempting to fit to a thermal profile naively requires almost constant-density, Solar-metallicity, uniformly hot (super-virial-temperature) gas out to the virial radius (in tension with limits from cosmology, the total metal budget of galaxies, and both UV and X-ray absorption-line studies). 

We show that the observed X-ray halos instead could naturally arise from inverse Compton (IC) scattering of the CMB by extended $\sim$\,GeV cosmic ray (CR) halos around galaxies.   Measurements of CRs in the ISM of the MW directly show that GeV CRs largely escape the galaxy LISM out into the CGM.  Simple models of steady-state CR profiles expected around galaxies (given the GeV CR density observed in the ISM) naturally predict the normalization, shape, and characteristic photon energy of the observed X-ray surface-brightness profiles. The CMB temperature and $\sim$\,GeV peak of the CR energy density ensure that the IC X-ray emission peaks in observed soft X-rays ($\sim 1\,$keV), with a spectrum that resembles hot gas thermal emission.  The total halo X-ray luminosity predicted from IC emission also agrees well with that observed around galaxies with stellar masses $M_{\ast}\lesssim 2\times 10^{11}\,M_{\odot}$.  The models predict associated synchrotron and $\gamma$-ray halos but we show that these would be extremely faint, well below present detection limits except in the MW and M31 specifically, where the predicted $\gamma$-ray emission assuming a CR-pressure dominated halo with LISM-like proton-to-electron ratio agrees well with Fermi observations. 

Our calculations assume that the local ISM CR electron spectrum observed in the MW is representative of the CR spectra in other galaxies with similar total stellar masses; if correct, the surprisingly quasi-thermal spectrum of IC-scattered CMB radiation should be generic in galaxy halos.  The assumption of a MW ISM CR electron spectrum is less easily justified in massive galaxies whose CRs likely originate primarily from AGN rather than star formation. 

We stress that CR-IC halos {\em must} be present at some level around MW-mass halos, given what is known about the energy density, lifetimes, and residence times of GeV CRs in the LISM ($\sim 8$\,kpc from the Galactic center), which necessarily imply that most of the GeV CR electron energy in the LISM leaks into the CGM.   Without invoking unphysically large CGM magnetic fields or gas densities (which are immediately ruled out by existing radio and $\gamma$-ray observations, and Galactic CR studies like those in \citealt{strong:2010.galprop.multiwavelength.cr.emission.across.wide.range,evoli:2019.cr.fitting.galprop.update.ams02}), the only way to strongly modify the IC halos would be to either (a) make the CR streaming speed much lower, which would result in the same halo {\em luminosity}, but a much brighter, more compact halo, which is clearly ruled out by the observations, or (b) make the streaming speed and diffusivity orders-of-magnitude larger (much larger than any reasonable physical model prediction), so CRs escape the halo much more rapidly and have low energy densities in the CGM.

We predict that CR-IC halos should dominate the X-ray emission for halos less massive than sizeable groups, above which the larger masses and hotter virial temperatures produce larger thermal emission. This appears to be supported by the keV halo luminosity $L_{X}$ versus stellar-mass $M_{\ast}$ correlation, which steepens dramatically above a stellar mass $\sim 1.5-2\times10^{11}\,M_{\odot}$. 
It is plausible, however, that that CR-IC could still contribute significantly to soft X-ray emission at radii $\lesssim 100-200\,$kpc even at these higher masses (well within the virial radii for the expected large halo masses, so not dominating the total luminosity), especially if strong AGN have been active in the last $\sim$\,Gyr, as argued for e.g.\ some ``mini-halos'' in massive clusters like Ophiuchus or Abell 3112   \citep{bartels:2015.radio.inverse.compton.cluster.minihalo.prospects,gitti:2002.perseus.minihalo.xray.inverse.compton,gitti:2004.minihalo.abell.2626.reaccel,bonamente:2007.abell.3112.clear.xray.inverse.compton.required.luminosity.fits.models.gamma.rays.too,murgia:2010.ophiuchus.cluster.minihalo.xray.inverse.compton,gitti:2016.radio.minihalos.coolcore.clusters.candidates.review}.

If CR-IC is indeed the source of the extended soft X-ray halos around low-mass galaxies, then the keV soft X-ray surface brightness is {\em directly} proportional to the CR lepton energy density, without any uncertain degenerate factors (unlike e.g.\ synchrotron and $\gamma$-ray emission, which are degenerate with magnetic field strengths and gas densities, respectively, and depend much more sensitively on the CR spectral shape).  This is because the seed CMB photons for IC scattering are precisely known.  The total halo X-ray luminosity (limited by IC losses) constrains the leptonic energy injection rate from the galaxy; observed values are consistent with expected SNe or AGN (if a fraction of just a few $10^{-5}$ of $\dot{M}_{\rm BH}\,c^{2}$ accelerates leptons) injection rates in MW and M31 and lower-mass systems. The X-ray surface brightness of the IC halos also yields a constraint on the effective diffusion coefficient (or streaming speed) of $\sim$\,GeV (rigidity $\sim$\,GV) CRs, which we show agrees well with ISM constraints close to galaxies, but $\kappa_{\rm eff}$ rises by an order of magnitude or more at $\gtrsim 100\,$kpc, consistent with a CR streaming speed of $\sim 100\,{\rm km\,s^{-1}}$, in agreement for MW-mass halos with independent (indirect) constraints from UV absorption studies \citep{butsky:2022.cr.kappa.lower.limits.cgm}. Assuming ISM-like proton-to-electron ratios, our results also imply that the CR pressure is a major component of the total gas pressure in the CGM of MW-mass galaxies, as suggested by
previous cosmological zoom-in simulations with CRs (e.g., \citealt{ji:fire.cr.cgm}).

Further characterizing the soft X-ray halos of MW-mass galaxies therefore presents a potentially invaluable probe of both CR and CGM physics. In future work, more detailed modeling of CR emission processes and X-ray spectra would be useful, to make more detailed quantitative predictions for future low-surface-brightness X-ray imaging and microcalorimetry experiments. Because the details of the CR-IC halo profiles depend on the behavior of CR scattering rates/diffusivities as a function of galacto-centric radius, this is a powerful probe of more detailed future models for the unknown, au-scale CR microphysics in the CGM, which will demand more detailed models to test directly against observations.

\begin{acknowledgements}
Support for PFH \&\ EQ was provided by Simons Investigator grants. PFH also acknowledges the support of the Institute for Advanced Study where early discussions leading to this work occurred. Numerical calculations were run on allocation AST21010 supported by the NSF.  This work was supported in part by NSF AST grant 2107872.
\end{acknowledgements}

\bibliographystyle{mn2e}
\bibliography{ms_extracted}

\end{document}